\input harvmac.tex
%\draftmode

%%%%%%%%%%%%%%%%%  THIS REFS ARE IN THE ORDER THEY COME OUT %%%%%%%%%%

\nref\dixon{L.J.~Dixon, ``Some World Sheet Properties of
Superstring Compactifications, on Orbifolds and Otherwise'', {\it
Lectures given at the 1987 ICTP Summer Workshop in High Energy
Phsyics and Cosmology, Trieste, Italy, Jun 29 - Aug 7, 1987}.}
\nref\ftheory{C.~Vafa, ``Evidence for
F-Theory", Nucl. Phys. {\bf B496} (1996) 403.}
\nref\GVW{S.~Gukov, C.~Vafa and E.~Witten, ``CFT's From
Calabi-Yau Four-folds", Nucl.\ Phys.\ {\bf B584} (2000) 69,
hep-th/9906070.}
\nref\Gukov{S.~Gukov, ``Solitons, Superpotentials and
Calibrations", Nucl.\ Phys.\ {\bf B574} (2000) 169,
hep-th/9911011.} 

%--------------+--- single ref. ---------+------------+----------+%

\nref\strom{A.\ Strominger,  ``Superstrings With Torsion'',
Nucl.\ Phys.\ {\bf B274} (1986) 253.}
\nref\RohmW{R.~Rohm and E.~Witten, ``The Antisymmetric Tensor Field in Superstring
  Theory'', Ann.\ Phys.\ {\bf 170} (1986) 454.}
\nref\DSHD{B.\ de Wit, D.J.\ Smit and N.D.\ Hari Dass, 
``Residual Supersymmetry of Compactified d = 10 Supergravity'',
Nucl.\ Phys.\ {\bf B283} (1987) 165.}
\nref\PS{J.~Polchinski and A.~Strominger, ``New Vacua for
Type II String Theory'', Phys. Lett. {\bf B388} (1996) 736,
hep-th/9510227.}

\nref\EW{E.\ Witten, ``Strong Coupling Expansion of Calabi-Yau
Compactification'', Nucl.\ Phys.\ {\bf B471} (1996) 135, hep-th/9602070.}
\nref\Michelson{J.~Michelson, ``Compactifications of Type IIB Strings to
Four Dimensions with Nontrivial Classical Potential'',
Nucl.\ Phys.\ {\bf B495} (1997) 127, hep-th/9610151.} 
\nref\Ovrut{A.~Lukas, B.~Ovrut and D.~Waldram, ``Stabilizing Dilaton
and Moduli Vacua in String and M-Theory Cosmology,''
Nucl. Phys. {\bf B509} (1998) 169, hep-th/9611204.} 
\nref\NOY{H.P.\ Nilles, M.\ Olechowski and M.\ Yamaguchi,
``Supersymmetry Breaking and Soft Terms in M-Theory'', Phys.\ Lett.\ 
{\bf B415} (1997) 24, hep-th/9707143.}

\nref\AGNT{I.~Antoniadis, E.~Gava, K.~S. Narain and T.~R. Taylor, ``Duality in 
Superstring Compactifications with Magnetic Field Backgrounds'', Nucl.\
Phys.\ {\bf B511} (1998) 611, hep-th/9708075.}
\nref\LOW{A.\ Lukas, B.A.\ Ovrut and D.\ Waldram,
``On the Four-Dimensional Effective Action of Strongly Coupled 
Heterotic String Theory'',
Nucl.\ Phys.\ {\bf B532} (1998) 43, hep-th/9710208.}
\nref\LOSW{A.\ Lukas, B.A.\ Ovrut, K.S.\ Stelle and D.\ Waldram,
``Heterotic M-Theory in Five Dimensions'',
Nucl.\ Phys.\ {\bf B552} (1999) 246, hep-th/9806051.}

\nref\TV{T.~R.~Taylor and C.~Vafa, ``RR Flux on Calabi-Yau and Partial
Supersymmetry Breaking'', Phys.\ Lett.\ {\bf B474}, 130 (2000), hep-th/9912152.} 
\nref\BG{K.~Behrndt and S.~Gukov, ``Domain Walls and
Superpotentials from M-Theory on Calabi-Yau Three-Folds'',
Nucl. Phys. {\bf B580} (2000) 225, hep-th/0001082.}
\nref\Mayr{P.\ Mayr, ``On Supersymmetry Breaking in String Theory  and its
Realization in Brane Worlds'', Nucl.\ Phys.\ {\bf B593} (2001) 99, hep-th/0003198.}
\nref\STaylor{M.\ Spalinski and T.R.\ Taylor,  ``Branes and Fluxes
in $D = 5$ Calabi-Yau Compactifications of M Theory'',
Phys. Lett. {\bf B485} (2000) 263, hep-th/0004095.}

\nref\GranaP{M.~Grana and J.~Polchinski, ``Supersymmetric Three-Form Flux 
Perturbations on AdS(5)'', Phys.\ Rev.\ {\bf D63} (2001) 026001, hep-th/0009211.}
\nref\Gubser{S.S.~Gubser, ``Supersymmetry and F-Theory Realization of the 
Deformed Conifold with Three-Form Flux'', hep-th/0010010.}
\nref\GD{G.~Dall'Agata, ``Type {IIB} Supergravity Compactified on a Calabi-Yau
Manifold with {H}-Fluxes'', JHEP 0111:005 (2001), hep-th/0107264.}
\nref\CKrause{G.\ Curio and A.\ Krause,  ``Four Flux
and Warped Heterotic M Theory  Compactifications'', 
Nucl.\ Phys.\ {\bf B602} (2001) 172, hep-th/0012152.}
\nref\CurioSC{G.~Curio, A.~Klemm, D.~L\"ust and S.~Theisen,
``On the Vacuum Structure of Type II String Compactifications on Calabi-Yau
Spaces with H-Fluxes'', Nucl.\ Phys.\ {\bf B609} (2001) 3, hep-th/0012213.}

\nref\GKP{S.B.~ Giddings, S.~Kachru and J.~Polchinski, ``Hierarchies from
Fluxes in String Compactifications'', hep-th/0105097.} 
\nref\Curiotwo{G.~Curio, A.~Klemm, B.~K\"ors and D.~L\"ust,
``Fluxes in Heterotic and Type II String Compactifications'',
Nucl. Phys. {\bf B620} (2002) 237, hep-th/0106155.}
\nref\CurioK{G.~Curio and A.~Krause, ``G-fluxes and Non-Perturbative
Stabilisation of Heterotic M-Theory'', hep-th/0108220.}
\nref\LM{J.~Louis and A.~Micu, ``Heterotic String Theory with Background
Fluxes'', Nucl.\ Phys.\ {\bf B626} (2002) 26, hep-th/0110187.}
\nref\lust{ G. Curio, B. K\"ors and D. L\"ust, ``Fluxes and Branes in Type II
Vacua and M Theory Geometry with $G_2$ and $Spin(7)$ Holonomy'', hep-th/0111165.}
\nref\AJ{J.~Louis and A.~Micu, ``Type II Theories Compactified on
Calabi-Yau Threefolds in the Presence of Background Fluxes'', hep-th/0202168.}

%%%%%%%%%%%%%%%%%%%%%%%%%%%%%%%%%%%%%%%%%%%%%%%%%%%%%%%%%%%%%%%%%%%%%%%%%%%%%%%%%%

\lref\ASpence{B.~S.~Acharya and B.~Spence, ``Flux, Supersymmetry
and M Theory on 7-Manifolds,'' hep-th/0007213.}
\lref\AGdO{B.~Acharya, S.~Gukov and X.~de la Ossa,
``Supersymmetry, G-flux and Spin(7) Manifolds'', hep-th/0201227.}
\lref\lavpop{I.~Lavrinenko, H.~L\"u and  C.~Pope, ``From Topology to Dimensional
Reduction'',  Nucl.\ Phys.\ {\bf B492} (1997) 278, hep-th/9611134.}

\lref\Vafalect{C.~Vafa, ``Lectures on Strings and
Dualities," hep-th/9702201.}
\lref\thesis{S.~Gukov, Ph.D. Thesis, Princeton University, 2001.}
\lref\GGWappear{S.J.~Gates, Jr., S.~Gukov and E.~Witten, ``Two
Two-Dimensional Supergravity Theories from Calabi-Yau Four-Folds",
Nucl.\ Phys.\ {\bf B584} (2000) 109, hep-th/0005120.}
\lref\ROM{ L.J.\ Romans, ``Massive $N=2A$
Supergravity in Ten Dimensions", Phys.\ Lett.\ {\bf B169} (1986) 374.}

\lref\BB{K.~Becker and M.~Becker, ``M-Theory on
Eight-Manifolds", Nucl.\ Phys.\ {\bf B477} (1996) 155, hep-th/9605053.}
\lref\BBtwo{K.~Becker and M.~Becker, ``Supersymmetry Breaking,
M Theory and Fluxes", JHEP 0107:038 (2001), hep-th/0107044.}
\lref\HaackL{M.~Haack and J.~Louis, ``M-theory Compactified
on Calabi-Yau Fourfolds With Background
Flux", Phys.\ Lett.\ {\bf B507} (2001) 296, hep-th/0103068.}
\lref\HLouis{M.~Haack and J.~Louis, ``Duality in Heterotic Vacua
With Four Supercharges", Nucl.\ Phys.\ {\bf B575} (2000) 107,
hep-th/9912181.}

\lref\MHLouis{M.~Haack, J.~Louis and M.~
Marquart, ``Type IIA and Heterotic String Vacua in D=2",
hep-th/0011075.}
\lref\DRS{K.~Dasgupta, G.~Rajesh and S.~Sethi,
``M-Theory, Orientifolds and G-Flux", JHEP 9908:023 (1999),
hep-th/9908088.}
\lref\deight{E.~Bergshoeff, M.~de Roo,
M.~B.~Green, G.~Papadopoulos and P.~K.~Townsend, ``Duality of Type
II 7-branes and 8-branes", Nucl.\ Phys.\ {\bf B470} (1996) 113.}
\lref\WittenBN{E.~Witten, ``Non-Perturbative Superpotentials in
String Theory,'' Nucl.\ Phys.\ {\bf B474} (1996) 343,
hep-th/9604030.}
\lref\GWehlau{M.T.~Grisaru and M.E.~Wehlau,
``Prepotentials for (2,2) Supergravity", Int.\ J.\ Mod.\ Phys.\
{\bf A10} (1995) 753, hep-th/9409043.}

\lref\GWmeasure{M.T.~Grisaru and
M.E.~Wehlau, ``Superspace Measures, Invariant Actions, and
Component Projection Formulae for $(2,2)$ Supergravity", Nucl.\
Phys.\ {\bf B457} (1995) 219, hep-th/9508139.}
\lref\GHR{S.J.~Gates, Jr., C.M.~Hull and M.~Rocek, ``Twisted
Multiplets and New Supersymmetric Non-linear Sigma-Models", Nucl.\
Phys.\ {\bf B248} (1984) 157.}
\lref\GGW{S.J.~Gates, Jr.,
M.T.~Grisaru and M.E.~Wehlau, ``A Study of General 2D, N=2 Matter
Coupled to Supergravity in Superspace´´, Nucl.\ Phys.\ {\bf B460}
(1996) 579, hep-th/9509021.}
\lref\VW{ C.\ Vafa and E.\ Witten, ``A One Loop Test
of String Duality", Nucl.\ Phys.\ {\bf B447} (1995) 261,
hep-th/9505053.}
\lref\DLM{ M.J.\ Duff, J.T.\ Liu and R.\
Minasian, ``Eleven-Dimensional Origin of String-String Duality: A
One Loop Test", Nucl.\ Phys.\ {\bf B452} (1995) 261,
hep-th/9506126.}

\lref\SVW{S.~Sethi, C.~Vafa and E.~Witten,
``Constraints on Low-Dimensional String Compactification'', Nucl.\
Phys.\ {\bf B480} (1996) 213, hep-th/9606122.}
\lref\DM{K.~Dasgupta and S.~Mukhi, ``A Note on Low-Dimensional
String Compactifications'',
 Phys.\ Lett.\ {\bf B398} (1997) 285, hep-th/9612188.}
\lref\GW{D.~Gross and E.~Witten, ``Superstring Modification of
Einstein Equations'', Nucl.\ Phys.\ {\bf B277} (1986) 1.}
\lref\FPSS{ M.D.\ Freeman, C.N.\ Pope, M.F.\ Sohnius and K.S.\
Stelle, ``Higher Order Sigma-Model Counterterms and the Effective
Action for Superstrings'', Phys.\ Lett.\ {\bf B178} (1986) 199.}
\lref\GVH{M.B.~Green and P.~Vanhove, ``D-Instantons, Strings and
M-Theory", Phys.\ Lett.\ {\bf B408} (1997) 122, hep-th/9704145.}

\lref\NEWSUSY{ E.~Bergshoeff, R.~Kallosh, T.~Ortin, D.~Roest and
A.~Van Proeyen, ``New Formulations of D = 10 Supersymmetry and D8
- O8 Domain Walls'', Class.\ Quant.\ Grav.\ {\bf 18} (2001) 3359,
hep-th/0103233.}
\lref\AT{A.A.~Tseytlin, ``$R^4$ Terms in $11$
Dimensions and Conformal Anomaly of $(2,0)$ Theory'', Nucl.\
Phys.\ {\bf B584} (2000) 233, hep-th/0005072.}
\lref\AS{A.~Strominger, ``Special Geometry'',
 Comm.\ Math.\ Phys.\ {\bf 133} (1990) 163.}
\lref\GS{M.B.~Green and J.H.~Schwarz, ``Supersymmetrical Dual
String Theory 2. Vertices and Trees", Nucl.\ Phys.\ {\bf B198}
(1982) 252.}
\lref\GVZ{M.T.~Grisaru, A.E.M.\ van de Ven and D.\
Zanon, ``Four-Loop $\beta$-Function for the $N=1$ and $N=2$
Supersymmetric Non-Linear Sigma Model in Two Dimensions", Phys.
Lett. {\bf B173} (1986) 423; ``Four Loop Divergences for the $N=1$
Supersymmetric Nonlinear Sigma Model in Two Dimensions", Nucl.\
Phys.\ {\bf B277} (1986) 409.}

\lref\GZ{M.T.~Grisaru and D.~Zanon,
``Sigma Model Superstring Corrections to the Einstein-Hilbert
Action", Phys.\ Lett.\ {\bf B177} (1986) 347.}
\lref\PZ{Q.~Park and D.~Zanon, ``More on Sigma Model Beta Functions
and Low-Energy Effective Actions", Phys.\ Rev.\ {\bf D35} (1987) 4038.}
\lref\Schwarz{J.H.~Schwarz, ``Superstring Theory", Phys.\ Rept.\
{\bf 89} (1982) 223.}
\lref\ST{ N.\ Sakai and Y.\ Tanii,
``One-Loop Amplitudes and Effective Action in Superstring
Theories", Nucl.\ Phys.\ {\bf B287} (1987) 457.}
\lref\FT{S.~Frolov and A.A.~Tseytlin, ``$R^4$ Corrections to the
Conifold and $G_2$-Holonomy Spaces", hep-th/0111128.}
\lref\AFMN{I.~Antoniadis, S.~Ferrara, R.~Minasian and K.S.~Narain,
``$R^4$ Couplings in M and Type II Theories on Calabi-Yau Spaces",
Nucl.\ Phys.\ {\bf B507} (1997) 571, hep-th/9707013.}
\lref\KP{E.~Kiritsis and B.~Pioline, ``On $R^4$ Threshold
Corrections in Type IIB String Theory and (p,q) String
Instantons", Nucl.\ Phys.\ {\bf B508} (1997) 509, hep-th/9707018.}

\lref\RT{J.G.~Russo and A.A.~Tseytlin, ``One-Loop Four Graviton
Amplitude in Eleven Dimensional Supergravity", Nucl.\ Phys.\ {\bf
B578} (2000) 139, hep-th/9707134.}
\lref\wvar{E.~Witten, ``String
Theory Dynamics in Various Dimensions", Nucl.\ Phys.\ {\bf B443}
(1995) 85, hep-th/9503124.}
\lref\FS{S.~Ferrara and S.~Sabharwal,
``Quaternionic Manifolds for Type II Superstring Vacua of
Calabi-Yau Spaces'', Nucl.\ Phys.\ {\bf B332} (1990) 317.}

\lref\Pmayr{P.~Mayr, ``Mirror Symmetry, N = 1 Superpotentials and 
Tensionless Strings on  Calabi-Yau Four-Folds'', Nucl.\ Phys.\ {\bf B494} (1997) 489,
hep-th/9610162.}
\lref\LercheZB{W.~Lerche, ``Fayet-Iliopoulos Potentials from
Fourfolds'', JHEP 9711:004 (1997), hep-th/9709146.}
\lref\Kaste{P.\ Kaste, ``On the Twisted Chiral Potential in 
2d and the Analogue of Rigid Special Geometry for 4-Folds'', JHEP 9906:021 (1999),
hep-th/9904218.}
\lref\bgv{N.~Berkovits, S.~Gukov and B.C.~Vallilo, 
``Superstrings in 2D Backgrounds with R-R Flux and New Extremal Black Holes'',
Nucl.\ Phys.\ {\bf B614} (2001) 195, hep-th/0107140.}

\lref\DuffRS{M.~J.~Duff, R.~Minasian and E.~Witten, ``Evidence for
Heterotic/Heterotic Duality'', Nucl.\ Phys.\ {\bf B465} (1996)
413, hep-th/9601036.}

\lref\DixonBG{L.~J.~Dixon, ``Some World Sheet Properties of
Superstring Compactifications, on Orbifolds and Otherwise'',
PUPT-1074 {\it Lectures given at the 1987 ICTP Summer Workshop in
High Energy Phsyics and Cosmology, Trieste, Italy, Jun 29 - Aug 7, 1987}.}
\lref\GSS{B.R.~Greene, K.~Schalm and G.~Shiu, ``Warped
Compactifications in M and F Theory'', Nucl.\ Phys.\ {\bf B584}
(2000) 480, hep-th/0004103.}
\lref\FKS{S.~Ferrara, R.~Kallosh and A.~Strominger, ``N=2 Extremal
Black Holes", Phys.\ Rev.\ {\bf D52} (1995) 5412, hep-th/9508072.} 
\lref\FK{S.~Ferrara and R.~Kallosh, ``Supersymmetry and Attractors'',
Phys.\ Rev.\ {\bf D54} (1996) 1514, hep-th/9602136;
``Universality of Supersymmetric Attractors'', Phys.\ Rev.\ {\bf D54} (1996)
1525, hep-th/9603090.}
\lref\FGK{S.~Ferrara, G.W.~Gibbons and R.~Kallosh,
``Black Holes and Critical Points in Moduli Space", Nucl.\ Phys.\
{\bf B500} (1997) 75, hep-th/9702103.}

\lref\Moore{G.~Moore, ``Arithmetic and
Attractors", hep-th/9807087.}
\lref\Wfb{E. Witten, ``Five-Brane
effective action in M theory'', J.Geom.Phys. {\bf 22} (1997) 103,
hep-th/9610234.}
\lref\BW{C.~Beasley and
E.~Witten, ``A Note on Fluxes and Superpotentials in M-Theory
Compactifications on Manifolds of $G_2$ Holonomy'',
hep-th/0203061.}
\lref\FP{A.R.~Frey and J.~Polchinski, ``${\cal N}
= 3$ Warped Compactifications'', hep-th/0201029.}
\lref\cyk{http://hep.itp.tuwien.ac.at/$\sim$kreuzer/CY/}
\lref\cys{http://www.th.physik.uni-bonn.de/th/Supplements/cy.html}

%%%%%%%%%%%%%%%%%%%%%  DE SITTER REFERENCES %%%%%%%%%%%%%%%%%%%%%%%

\lref\SSV{M.~Spradlin, A.~Strominger and A.~Volovich, ``Les Houches
Lectures on De Sitter Space'', hep-th/0110007.}
\lref\Silverstein{E.~Silverstein, ``(A)dS Backgrounds from Asymmetric
Orbifolds'', hep-th/0106209.}
\lref\MNunez{J.~Maldacena and C.~Nunez, ``Supergravity Description
of Field Theories on Curved Manifolds and a No Go Theorem'',
Int.\ J.\ Mod.\ Phys.\ {\bf A16} (2001) 822, hep-th/0007018.}
\lref\KachruS{S.~Kachru, M.~Schulz and S.~Trivedi, ``Moduli Stabilization from Fluxes
in a Simple IIB Orientifold,'' hep-th/0201028.}
\lref\KPV{S.~Kachru, J.~Pearson and H.~Verlinde, ``Brane/Flux Annihilation
and String Dual of a Nonsupersymmetric Field Theory'', hep-th/0112197.}
\lref\BerglundAJ{P.~Berglund, T.~H\"ubsch and D.~Minic,
``De Sitter Spacetimes from Warped Compacti\-fications of IIB String  Theory'',
hep-th/0112079.}
\lref\BHM{V.~Balasubramanian, P.~Horava and D.~Minic,
``Deconstructing de Sitter'', JHEP 0105:043 (2001), hep-th/0103171.}

\lref\GibbonsWY{G.~W.~Gibbons and C.M.~Hull,
``De Sitter Space from Warped Supergravity Solutions'',
hep-th/0111072.}
\lref\HullYG{C.M.~Hull,
``Domain Wall and de Sitter Solutions of Gauged Supergravity'', 
JHEP 0111:061 (2001), hep-th/0110048.}
\lref\Hull{C.M.~Hull,
``Timelike T-Duality, de Sitter Space, Large N Gauge Theories and
Topological Field Theory'', JHEP 9807:021 (1998), hep-th/9806146;
``De Sitter Space in Supergravity and M Theory'',
JHEP 0111:012 (2001), hep-th/0109213.}
\lref\ChamblinL{ A.~Chamblin and N.D.~Lambert,
``De Sitter Space from M-Theory'', Phys.\ Lett.\ B {\bf 508} (2001) 369, 
hep-th/0102159;
``Zero-Branes, Quantum Mechanics and the Cosmological Constant'',
Phys.\ Rev.\ D {\bf 65} (2002) 066002, hep-th/0107031.}

%%%%%%%%%%%%%%% G2 REFERENCES %%%%%%%%%%%%%%%%%%%%%%%%%%%%%%%%%%%%%%

\lref\gubser{S.S.~Gubser, ``TASI lectures: Special Holonomy in
String Theory and M-theory'', hep-th/0201114.} \lref\Bobby{
B.S.~Acharya, ``On Realising N=1 Super Yang-Mills in M theory,''
hep-th/0011089.} \lref\AMV{ M.~Atiyah, J.~Maldacena and C.~Vafa,
``An M-theory flop as a large N duality,'' {{ hep-th/0011256}}.}

\lref\AW{ M.F.~Atiyah, E.~Witten, ``M-Theory Dynamics On A
Manifold of $G_2$ Holonomy,'' {{ hep-th/0107177}}.} \lref\Acharya{
B.~S.~Acharya, ``Confining strings from G(2)-holonomy
spacetimes,'' arXiv:hep-th/0101206.} \lref\AcharyaV{ B.~S.~Acharya
and C.~Vafa, ``On domain walls of N = 1 supersymmetric Yang-Mills
in four dimensions,'' arXiv:hep-th/0103011.} \lref\Partouche{
H.~Partouche and B.~Pioline, ``Rolling among G(2) vacua,'' JHEP
{\bf 0103}, 005 (2001) [arXiv:hep-th/0011130].} \lref\KasteKP{
P.~Kaste, A.~Kehagias and H.~Partouche, ``Phases of supersymmetric
gauge theories from M-theory on G(2)  manifolds,'' JHEP {\bf
0105}, 058 (2001) [arXiv:hep-th/0104124].} \lref\AganagicV{
M.~Aganagic and C.~Vafa, ``Mirror symmetry and a G(2) flop,''
arXiv:hep-th/0105225.} \lref\Brandhuber{ A.~Brandhuber, J.~Gomis,
S.~S.~Gubser and S.~Gukov, ``Gauge theory at large N and new G(2)
holonomy metrics,'' Nucl.\ Phys.\ B {\bf 611}, 179 (2001)
[arXiv:hep-th/0106034].} \lref\Wittenanomaly{ E.~Witten, ``Anomaly
cancellation on G(2)manifolds,'' arXiv:hep-th/0108165.}

\lref\Sugiyama{ K.~Sugiyama and S.~Yamaguchi, ``Cascade of special
holonomy manifolds and heterotic string theory,''
arXiv:hep-th/0108219.} \lref\Blumenhagen{ R.~Blumenhagen and
V.~Braun, ``Superconformal field theories for compact G(2)
manifolds,'' arXiv:hep-th/0110232.} \lref\Eguchi{ T.~Eguchi and
Y.~Sugawara, ``String theory on G(2) manifolds based on Gepner
construction,'' arXiv:hep-th/0111012.} \lref\Ferretti{
G.~Ferretti, P.~Salomonson and D.~Tsimpis, ``D-brane probes on
G(2) orbifolds,'' arXiv:hep-th/0111050.} \lref\Blumenhagen{
R.~Blumenhagen and V.~Braun, ``Superconformal field theories for
compact manifolds with Spin(7)  holonomy,'' arXiv:hep-th/0111048.}
\lref\AWchiral{ B.~Acharya, E.~Witten, ``Chiral Fermions from
Manifolds of $G_2$ Holonomy,'' hep-th/0109152.}
\lref\decon{ E. Witten, ``Deconstruction, $G_2$
Holonomy and Doublet-Triplet Splitting,'' hep-ph/0201018.}

%%%%%%%%%%%%%% END OF G2 REFERENCES %%%%%%%%%%%%%%%%%%%%%%%%%%%%%

%%%%%%%%%%%%%%%%%%%%%  Math-style letters   %%%%%%%%%%%%%%%%%%%%%%%%
\font\cmss=cmss10 \font\cmsss=cmss10 at 7pt

\def\IB{\relax\hbox{$\inbar\kern-.3em{\rm B}$}}
\def\IC{\relax\hbox{$\inbar\kern-.3em{\rm C}$}}
\def\ID{\relax\hbox{$\inbar\kern-.3em{\rm D}$}}
\def\IE{\relax\hbox{$\inbar\kern-.3em{\rm E}$}}
\def\IF{\relax\hbox{$\inbar\kern-.3em{\rm F}$}}
\def\IG{\relax\hbox{$\inbar\kern-.3em{\rm G}$}}
\def\IGa{\relax\hbox{${\rm I}\kern-.18em\Gamma$}}
\def\IH{\relax{\rm I\kern-.18em H}}
\def\IK{\relax{\rm I\kern-.18em K}}
\def\IL{\relax{\rm I\kern-.18em L}}
\def\IP{\relax{\rm I\kern-.18em P}}
\def\IR{\relax{\rm I\kern-.18em R}}
\def\Z{\relax\ifmmode\mathchoice
{\hbox{\cmss Z\kern-.4em Z}}{\hbox{\cmss Z\kern-.4em Z}}
{\lower.9pt\hbox{\cmsss Z\kern-.4em Z}}
{\lower1.2pt\hbox{\cmsss Z\kern-.4em Z}}\else{\cmss Z\kern-.4em Z}\fi}
\def\IZ{Z\!\!\!Z}
\def\II{\relax{\rm I\kern-.18em I}}

%%%%%%%%%%%%%%%%%%%%% Calligraphic letters  %%%%%%%%%%%%%%%%%%%%%

\def\CE {{\cal E}}

\def\CK {{\cal K}}
\def\CL {{\cal L}}

\def\CN {{\cal N}}

\def\CV {{\cal V}}

\def\CX {{\cal X}}
\def\CY {{\cal Y}}

%%%%%%%%%%%%%%%%%%%%%%%%%% Derivatives  %%%%%%%%%%%%%%%%%%%%%%%%

%%%%%%%%%%%%%%%%%%%% letters with bar %%%%%%%%%%%%%%%%%%%%%%%%%%
\def\tilde{\widetilde}
\def\hat{\widehat}
\def\bar{\overline}

\def\nb {{\bar{\nabla}}}

%%%%%%%%%%%%%%%%%%%%%%%%%%% Math symbols %%%%%%%%%%%%%%%%%%%%%%%

\def\vol{{\rm vol}}

\def\inbar{\,\vrule height1.5ex width.4pt depth0pt}

%%%%%%%%%%%%%%%%%%%   Greek letters %%%%%%%%%%%%%%%%%%%
\def\a{\alpha}

\def\la{\lambda}
\def\th{\theta}

%%%%%%%%%%%%  Jim Gates and Marc Grisaru MACROS   %%%%%%%%%%%%%%%
\def\pp{{\mathchoice
            %{general format
               %[w] = length of horizontal bars
               %[t] = thickness of the lines
               %[h] = length of the vertical line
               %[s] = spacing around the symbol
              %
              %\kern [s] pt%
              %\raise 1pt
              %\vbox{\hrule width [w] pt height [t] pt depth0pt
              %      \kern -([h]/3) pt
              %      \hbox{\kern ([w]-[t])/2 pt
              %            \vrule width [t] pt height [h] pt depth0pt
              %            }
              %      \kern -([h]/3) pt
              %      \hrule width [w] pt height [t] pt depth0pt}%
              %      \kern [s] pt
          {%displaystyle
              \kern 1pt%
              \raise 1pt
              \vbox{\hrule width5pt height0.4pt depth0pt
                    \kern -2pt
                    \hbox{\kern 2.3pt
                          \vrule width0.4pt height6pt depth0pt
                          }
                    \kern -2pt
                    \hrule width5pt height0.4pt depth0pt}%
                    \kern 1pt
           }
            {%textstyle
              \kern 1pt%
              \raise 1pt
              \vbox{\hrule width4.3pt height0.4pt depth0pt
                    \kern -1.8pt
                    \hbox{\kern 1.95pt
                          \vrule width0.4pt height5.4pt depth0pt
                          }
                    \kern -1.8pt
                    \hrule width4.3pt height0.4pt depth0pt}%
                    \kern 1pt
            }
            {%scriptstyle
              \kern 0.5pt%
              \raise 1pt
              \vbox{\hrule width4.0pt height0.3pt depth0pt
                    \kern -1.9pt  %[e]=0.15pt
                    \hbox{\kern 1.85pt
                          \vrule width0.3pt height5.7pt depth0pt
                          }
                    \kern -1.9pt
                    \hrule width4.0pt height0.3pt depth0pt}%
                    \kern 0.5pt
            }
            {%scriptscriptstyle
              \kern 0.5pt%
              \raise 1pt
              \vbox{\hrule width3.6pt height0.3pt depth0pt
                    \kern -1.5pt
                    \hbox{\kern 1.65pt
                          \vrule width0.3pt height4.5pt depth0pt
                          }
                    \kern -1.5pt
                    \hrule width3.6pt height0.3pt depth0pt}%
                    \kern 0.5pt%}
            }
        }}

\def\mm{{\mathchoice
                      %{general format %[w] = length of bars
                                       %[t] = thickness of bars
                                       %[g] = gap between bars
                                       %[s] = space around symbol
   %[w], [t], [s], [h]=3([g]) are taken from corresponding definitions of \pp
   %
                      %       \kern [s] pt
               %\raise 1pt    \vbox{\hrule width [w] pt height [t] pt depth0pt
               %                   \kern [g] pt
               %                   \hrule width [w] pt height[t] depth0pt}
               %              \kern [s] pt}
                  %
                       {%displaystyle
                             \kern 1pt
               \raise 1pt    \vbox{\hrule width5pt height0.4pt depth0pt
                                  \kern 2pt
                                  \hrule width5pt height0.4pt depth0pt}
                             \kern 1pt}
                       {%textstyle
                            \kern 1pt
               \raise 1pt \vbox{\hrule width4.3pt height0.4pt depth0pt
                                  \kern 1.8pt
                                  \hrule width4.3pt height0.4pt depth0pt}
                             \kern 1pt}
                       {%scriptstyle
                            \kern 0.5pt
               \raise 1pt
                            \vbox{\hrule width4.0pt height0.3pt depth0pt
                                  \kern 1.9pt
                                  \hrule width4.0pt height0.3pt depth0pt}
                            \kern 1pt}
                       {%scriptscriptstyle
                           \kern 0.5pt
             \raise 1pt  \vbox{\hrule width3.6pt height0.3pt depth0pt
                                  \kern 1.5pt
                                  \hrule width3.6pt height0.3pt depth0pt}
                           \kern 0.5pt}
                       }}

\def\ad{{\kern0.5pt
                   \alpha \kern-5.05pt
\raise5.8pt\hbox{$\textstyle.$}\kern 0.5pt}}

\def\bd{{\kern0.5pt
                   \beta \kern-5.05pt \raise5.8pt\hbox{$\textstyle.$}\kern 0.5pt}}

\def\qd{{\kern0.5pt
                   q \kern-5.05pt \raise5.8pt\hbox{$\textstyle.$}\kern 0.5pt}}
\def\Dot#1{{\kern0.5pt
     {#1} \kern-5.05pt \raise5.8pt\hbox{$\textstyle.$}\kern 0.5pt}}
%%%%%%%%%%%%%%%%%%%%%%%%%%%%%%%%%%%%%%%%%%%%%%%%%

\Title{\vbox{\baselineskip12pt\hbox{hep-th/0203267}
\hbox{ROM2F/2002/07} \hbox{ITEP-TH-07/02} \hbox{HUTP-01/A074}}} {\vbox{ \centerline{IIA
String Theory on Calabi-Yau Fourfolds} \centerline{with Background
Fluxes} \vskip 4pt
%\centerline{}
}}
\centerline{Sergei Gukov$^{\clubsuit}$ and Michael Haack$^{\spadesuit}$}
%\medskip
%\medskip
%\medskip
\medskip
\vskip 8pt
\centerline{\it $^{\clubsuit}$ Jefferson Physical Laboratory}
\centerline{\it Harvard University}
\centerline{\it Cambridge, MA 02138, USA}
\medskip
\medskip
\centerline{\it $^{\spadesuit}$ Dipartimento di Fisica}
\centerline{\it Universit\`a di Roma, ``Tor Vergata"}
\centerline{\it 00133 Rome, Italy}
\medskip
\medskip
%\medskip
\medskip
%\bigskip
\noindent

Looking for string vacua with fixed moduli,
we study compactifications of type IIA string theory on Calabi-Yau
fourfolds in the presence of generic Ramond-Ramond fields.
We explicitly derive the (super)potential induced by
Ramond-Ramond fluxes performing a Kaluza-Klein reduction
of the ten-dimensional effective action. This can be conveniently
achieved in a formulation of the massive type IIA supergravity
where all Ramond-Ramond fields appear in a democratic way.
The result agrees with the general formula for the superpotential
written in terms of calibrations.
We further notice that for generic Ramond-Ramond fluxes
all geometric moduli are stabilized and one finds
non-supersymmetric vacua at positive values of the
scalar potential.

\smallskip
\Date{March 2002}

%%%%%%%%%%%%%%%%%%%%%%%%%%%%%%%%%%%%%%%%%%%%%%%%%%%%%%%%%%%%%%%%%%%%%%%%%%%

\newsec{Introduction and Summary}

The minimal $\CN=1$ supersymmetry in $3+1$ dimensions
is a very attractive framework for addressing many outstanding
problems in theoretical high energy physics, including
unification of the gauge couplings, the hierarchy problem
and maybe even the cosmological constant problem.
Moreover, there are phenomenological indications for
minimal supersymmetry just above the electroweak scale,
in the energy range of about 100 GeV -- 1 TeV.
This strongly motivates the study of string models with
four supercharges.

For a long time, compactifications of the heterotic $E_8 \times E_8$
string theory on Calabi-Yau three-folds have been the leading candidates
for constructing realistic models with $\CN=1$ supersymmetry.
The other superstring theories were much less interesting;
there was even a no-go theorem proving the impossibility of obtaining
the Standard Model out of perturbative type II theories \dixon.
The situation has changed, however, with the discovery
of string dualities which allow to make definite statements
even in the regions of moduli space where perturbation theory cannot be used.
Thus phenomenologically interesting $\CN=1$ four-dimensional string vacua
can be equivalently described as non-perturbative vacua of type IIB string
theory known as F-theory compactifications on elliptically fibered
Calabi-Yau fourfolds \ftheory, or also as M-theory compactifications
on singular $G_2$ holonomy manifolds that recently received a lot of attention.

However, all attempts to get four-dimensional $\CN=1$ supergravity
out of string or M-theory face the so called moduli problem.
Typically there are a number of massless scalars in the spectrum
which arise e.g. due to geometric moduli of the
compactification-manifold. A possible solution to the moduli
problem lies in turning on background fluxes in the vacuum. These
induce effective superpotentials of the form \refs{\GVW,\Gukov}
\eqn\superw{ W = \int_{X} {\rm Flux} \wedge {\rm Calibration}\ ,}
with $X$ the compactification-manifold, and therefore generically
lift at least part of the moduli space. Due to this
phenomenological prospect type II and M-theory compactifica\-tions
on Calabi-Yau threefolds with fluxes and their heterotic
counterparts have attracted a lot attention in the past \strom\ -- \AJ.
%\refs{\strom,\RohmW,\DSHD,\PS,\EW,\Michelson,\Ovrut,\NOY,\AGNT,\LOW,
%\LOSW,\TV,\BG,\Mayr,\STaylor,\GranaP,\Gubser,\GD,\CKrause,\CurioSC,\GKP,
%\Curiotwo,\CurioK,\LM,\lust,\AJ}.
Also the potential induced by 4-form fluxes in
compacti\-fications of M-theory on $G_2$ and $Spin(7)$ manifolds
has been investigated \refs{\Gukov,\ASpence, \AGdO, \BW}. All these cases
have in common that the potential can indeed be expressed via a
superpotential of the form \superw\ and that it is difficult to
stabilize all the moduli.

Recently, Giddings, Kachru and Polchinski proposed an elegant way
to generate large hierarchies and to stabilize some of the moduli
in warped $\CN=1$ compactifications of F-theory on Calabi-Yau
fourfolds with background fluxes \GKP.
%The basic idea is that background fluxes induce effective
%superpotentials of the form \refs{\GVW,\Gukov}:
%
%\eqn\superw{ W = \int_{Y_4} {\rm Flux} \wedge {\rm Calibration} }
%
%and therefore generically lift most of the moduli fields,
%but not the overall volume.
% \refs{\CurioSC}.
However, in the case of F-theory compactifications with fluxes the
K\"ahler moduli and specifically the volume modulus remain unfixed
\GKP\ (see however \KachruS). The problem is expected to be resolved
by quantum corrections, which modify the classical expression
for the effective potential. 
%In particular, due to constraints from
%supersymmetry and holomorphy one would hope for a complete analytic
%description of the non-perturbative corrections to the superpotential.
In string theory, there are e.g.~non-perturbative corrections to the superpotential 
from five-brane instantons
\refs{\WittenBN,\thesis}, and it is natural to speculate that
competing effects of background fluxes and five-brane instantons
result in a superpotential that leads to isolated minima.

%\foot{However, one should keep in mind that a
%non-trivial 4-form flux through a 6-cycle of the Calabi-Yau
%fourfold prevents a fivebrane from wrapping this cycle \Wfb.}
%Unfortunately, it is hard to make this explicit
%due to the inherent non-perturbative nature of F-theory.

Closely related to compactifications of F-theory on Calabi-Yau
fourfolds are compacti\-fications of M-theory and type IIA string
theory on the same manifolds \ftheory. Albeit less interesting for
phenomenological applications, the resulting $\CN=2$
(resp.~$\CN=(2,2)$) supersymmetric theories in $2+1$ (resp.~$1+1$)
dimensions exhibit many similar features as $\CN=1$
four-dimensional vacua of F-theory. Yet, they are more tractable
because the low-energy physics of M-theory and IIA string theory
can be described in terms of effective supergravity theories,
which are suitable starting points for Kaluza-Klein reductions
with background fluxes.
%Although the expected form of the effective superpotential \superw\
%looks very simple and agrees with the analysis of the supersymmetry
%conditions in supergravity \refs{\GVW,\Gukov,\ASpence,\AGdO},
%it is usually hard to derive $W$ by direct Kaluza-Klein reduction.
%In the case of Calabi-Yau three-folds the problem becomes more tractable,
%and it was discussed in a number of recent papers
%\refs{\PS,\Michelson,\Ovrut,\TV,\BG,\CurioSC,\Curiotwo}.
%On the other hand,
The explicit compactification of M-theory on Calabi-Yau fourfolds
with background 4-form fluxes has been studied in \HaackL\ (see
also \refs{ \BB, \Pmayr, \GVW, \DRS, \GSS, \BBtwo}) with the result that
the potential can again be expressed via superpotentials of the
form \superw.
%However, although the K\"ahler moduli get a non-trivial
%potential in this case, the volume modulus still can not be fixed.
A similar problem of direct computation of the scalar potential
induced by 4-form flux in M-theory on $G_2$ holonomy manifold
has been addressed recently in \BW.
%. It was
%found that the resulting (super)potential induced by the flux
%again agrees with the expected expression \superw.

In this paper we address the corresponding problem for type IIA
compactifications on Calabi-Yau fourfolds (see also \refs{\Pmayr,
\LercheZB, \Kaste, \Gukov, \GGWappear, \MHLouis, \bgv}). By performing a 
direct Kaluza-Klein
reduction we explicitly derive the (super)potential induced by
generic Ramond-Ramond fluxes. As in the previous examples, we find
complete agreement with the general formula \superw. Since in type
IIA theory there are more possible fluxes, the analysis is more
complicated and requires a special care to treat all Ramond-Ramond
fields democratically \NEWSUSY. On the other hand, for the same
reason, the structure of the induced superpotential is richer. In
particular, it allows
%many possibilities to break supersymmetry and
to stabilize all geometric moduli of the Calabi-Yau fourfold including its volume.

The main result of the paper is the explicit formula for the scalar
potential (eqn. (3.23) below), which can be expressed in terms of
a superpotential $W$ depending on the complex structure moduli
and a twisted superpotential $\hat W$ depending on
the (complexified) K\"ahler moduli of the Calabi-Yau fourfold.
Both of them are of the form \superw, {\it cf}. sect. 3.
In terms of $W$ and $\hat W$ the potential can be expressed as:
%the (twisted) superpotential \superw\ in the usual way:
%
\eqn\vpreview{
%V = e^{K_{1,1}} \vert D \hat W \vert^2 + e^{K_{3,1}} \vert DW \vert^2
V =  e^{K_{3,1}} g^{-1 \alpha\bar \beta} D_\alpha W
D_{\bar \beta} \bar W + e^{K_{1,1}} g^{-1 A \bar B} D_A \hat W D_{\bar B}
\bar{\hat W}\ ,}
where $D_A$ and $D_\alpha$ are appropriate K\"ahler covariant derivatives
%
%\eqn\covd{ D_A W = \p_A W + (\p_A K_{3,1}) W,
%\quad D_{\a} \hat W = \p_{\a} \hat W + (\p_{\a} K_{1,1}) \hat W}
%
and $K_{3,1}$ (resp. $K_{1,1}$) is the K\"ahler potential
of the complex structure (resp. K\"ahler moduli). Note, that
the potential splits into one part for the complex structure
moduli and one for the K\"ahler moduli. Furthermore, similar to
the four-dimensional \GKP\ and three-dimensional \HaackL\ case,
eqn.~\vpreview\ does not contain any $|W|^2$ and $| \hat W |^2$ terms.

It would be interesting to embed our analysis in a manifestly
supersymmetric description of the effective two-dimensional theory.
A natural framework for this would be a superspace description
of the effective $\CN=(2,2)$ supergravity coupled to matter,
e.g. as in the hybrid description of the superstring \bgv\
or in the new two-dimensional dilaton supergravity \GGWappear.
Even though such a description can be obtained 
%this can be achieved
in the case without potential,
we were not able to incorporate the superpotentials
in $\CN=(2,2)$ superspace. Nevertheless, in the appendix we derive
the chiral density projectors in the ungauged $\CN=(2,2)$ dilaton
supergravity of \GGWappear, which might not only play a role
in solving this problem, but also be of independent interest
in the general study of two-dimensional supergravity
in superspace \refs{\GGWappear,\GGW,\GWehlau,\GWmeasure}.

The organization of the paper is as follows. In section 2, using
the dual formulation of \NEWSUSY , we derive a form of the
ten-dimensional massive type IIA theory, which breaks the
ten-dimensional covariance but serves as a convenient starting
point for a compactification on a Calabi-Yau fourfold with fluxes.
This compactification is done in section 3, where
the potential due to the Ramond-Ramond background fluxes is derived.
The derivation also takes into account some of
the higher derivative corrections to the type IIA action.
Furthermore we find a more general tadpole cancellation condition
than in three dimensions. At the end of this section we make some
comments about the vacuum structure of the two-dimensional theory.
Like in compactifications to three dimensions,  we find no
stable AdS or dS vacua. On the other hand, there are more
possibilities to break supersymmetry and stabilize moduli
than in $2+1$ and $3+1$ dimensions.
In section 4 we illustrate the rather technical
derivation in section 3 by looking at three cases, in which only
a single type of RR-field strength has a background value.
Specifically, we consider the cases of non-vanishing 0-, 8- and 4-form flux.
This should help the reader to understand the physical implications
of the different fluxes and the connection to the three-dimensional case,
in which only $4$-form flux can appear in the background.
Finally, we have included two appendices.
The first one collects some of our conventions and useful formulas.
In appendix II (which can be read independently of the rest of
the paper) we derive the chiral and twisted chiral
density projectors in $\CN=(2,2)$
two-dimensional dilaton supergravity.

%%%%%%%%%%%%%%%%%%%%%%%%%%%%%%%%%%%%%%%%%%%%%%%%%%%%%%%%%%%%%%%%%%

\newsec{A Dual Formulation of Massive Type IIA Theory}

In this section we work out a formulation of Type IIA supergravity
suitable for compactifications on 8-manifolds, which we will use
to perform a compactification on a  Calabi-Yau fourfold $Y_4$ in
the next section. The effective action we are going to obtain
breaks the full ten-dimensional covariance, and makes a $2+8$
split of the coordinates. However, it 
has the same degrees of freedom as the usual (massive)
Type IIA theory, albeit written in a different, dual language.

The standard bosonic action of the massive Type IIA supergravity
looks like \ROM :
\eqn\iiastand{\eqalign{ S_{10} & =  {1 \over 2
\kappa_{10}^2} \int d^{10} x \sqrt{-g} \Big[e^{-2 \varphi} (R + 4
|\partial \varphi|^2 - {1 \over 2} |H|^2 ) - {1 \over 2}
\sum_{n=0,1,2} |G_{2n}|^2\Big] - \cr & - {1 \over 4 \kappa_{10}^2}
\int \Big(dC_3 dC_3 B + {1 \over 3} G_0 dC_3 B^3 + {1 \over 20}
G^{2}_0 B^5 \Big) + \ldots\ , }}
where we omit the wedge products and use the abbreviations
\eqn\gdc{\eqalign{ G_0 & = {\rm
mass~parameter}\ , \cr G_2 & = dC_1 + G_0 B\ , \cr G_4 & = dC_3 -
C _1 d B + {1 \over 2} G_0 B^2\ . }}
Furthermore for a $p$-form
$\omega_{p}$ we use the notation $|\omega|^2 = {1 \over p!}
\omega_{M_1 \ldots M_p} \omega^{M_1 \ldots M_p}$ and the dots in
\iiastand\ stand for higher derivative terms.
This standard action can be obtained via integrating out the
fields $A_{(2n-1)}$ in the dual action \NEWSUSY:
\eqn\iiadual{\eqalign{ S_{10} & = {1 \over 2 \kappa_{10}^2} \int
d^{10} x \sqrt{-g} \Big[e^{-2 \varphi} (R + 4 |\partial \varphi|^2
- {1 \over 2} |H|^2 ) - {1 \over 2} \sum_{n=0,1,2} |G_{2n}|^2\Big]
- \cr & - {1 \over 4 \kappa_{10}^2} \int \Big(G_4 G_4 B - G_2 G_4
B^2 + {1 \over 3}  G^{2}_2 B^3 + {1 \over 3} G_0 G_4 B^3 - {1
\over 4} G_0 G_2 B^4 + \cr & + {1 \over 20} G^{2}_0 B^5 + 2 e^{-B}
{\bf G} d (A_5 - A_7 + A_9) \Big) + \ldots\ ,}}
where
\eqn\bfg{ {\bf G} = G_0 + G_2 + G_4}
is a formal sum of Ramond-Ramond
fields, which are however a priori independent fields, not given
by \gdc\ (compare the discussion in \NEWSUSY). Moreover, in the
last term of \iiadual\ a projection on the $10$-form part is
understood. The equations of motion for $A$ now impose the
following Bianchi identities for ${\bf G}$: \eqn\eomaa{ d(e^{-B}
\wedge {\bf G}) = 0\ . } This set of equations has the general
solution\foot{${\bf A} = A_1 + A_3$ is a formal sum like ${\bf G}$
in \bfg. Furthermore the $A$s are not necessarily globally
defined. In the presence of background fluxes considered in the
next section, $d{\bf A}$ is a nontrivial cohomology element.}
\eqn\solga{ e^{-B} \wedge {\bf G} = d {\bf A}\ .}
This can be expressed in components:
\eqn\solng{\eqalign{ G_0 & =
{\rm mass~parameter}\ , \cr G_2 & = dA_1 + G_0 B\ , \cr G_4 & =
dA_3 + G_2 B - {1 \over 2} G_0 B^2\ . }}
Converting these into the
``$C$-basis''
\eqn\aversuscbasis{ {\bf A} = {\bf C} \wedge e^{-B}}
and substituting back into \iiadual, we get the standard Type
IIA action \iiastand.

In principle it would also be possible to integrate out the fields
${\bf G}$ in \iiadual\ keeping $A_5, A_7$ and $A_9$. However, in
view of our aim to perform a Calabi-Yau fourfold
compacti\-fication with background fluxes, we want to proceed in a
different way, keeping some components of the $A$s and some of the
$G$s. To be more specific, we distinguish two of the coordinates
($x^\mu, \mu =1,2$) from the others ($x^a, a=1, \ldots, 8$),
denoting them as 'external' since they will be the uncompactified
coordinates in the $Y_4$-compactification in the next section. At
this stage this name helps us to distinguish them from the $x^a$.
What we have in mind now is to integrate out all components from
${\bf G}$ in \iiadual\ with two external indices, keeping those
with zero or one (and instead integrating out the corresponding
components of $A_5, A_7$ and $A_9$). This leads to an action which 
is (classically) equivalent to  \iiastand\ and has
the same degrees of freedom but expressed in
different variables, which are more adapted to a Calabi-Yau
fourfold compactification with background fluxes.

In order to carry out this program we have to introduce some
notation. For any $p$-form we define
\eqn\omegadecomp{\eqalign{
\omega_p^{(0)} & = {1 \over p!} \omega_{a_1 \ldots a_p} dx^{a_1}
\wedge \ldots \wedge dx^{a_p}\ , \cr \omega_p^{(1)} & = {1 \over
(p-1)!} \omega_{\mu a_1 \ldots a_{p-1}} dx^\mu \wedge dx^{a_1}
\wedge \ldots \wedge dx^{a_{p-1}}\ , \cr \omega_p^{(2)} & = {1
\over (p-2)!} {1 \over 2!} \omega_{\mu \nu a_1 \ldots a_{p-2}}
dx^\mu \wedge dx^\nu \wedge dx^{a_1} \wedge \ldots \wedge
dx^{a_{p-2}}\ , }}
so that
\eqn\omdecom{\eqalign{\omega_p =
\omega_p^{(0)} + \omega_p^{(1)} + \omega_p^{(2)}\ .}}
Furthermore, we define exterior derivatives
\eqn\ddecomp{\eqalign{d^{(0)}
\omega_p^{(0)} & = {1 \over p!} {\partial \over \partial x^b}
\omega_{a_1 \ldots a_p} dx^b \wedge dx^{a_1} \wedge \ldots \wedge
dx^{a_p}\ , \cr d^{(1)} \omega_p^{(0)} & = {1 \over p!} {\partial
\over \partial x^\mu} \omega_{a_1 \ldots a_p} dx^\mu \wedge
dx^{a_1} \wedge \ldots \wedge dx^{a_p}}}
and similarly for $\omega_p^{(1)}$ and
$\omega_p^{(2)}$, such that $d = d^{(0)} + d^{(1)}$.

Now we decompose all forms $G_2, G_4, A_5, A_7, A_9$ and $B$ as in
\omdecom . Before substituting these decompositions into the
action \iiadual, it is convenient to write the Lagrangian as a sum
\eqn\ldecomp{ L_{10} = T_{NS-NS} + T_{R-R} + L_{CS}\ ,}
where
\eqn\ttcs{\eqalign{ T_{NS-NS} & =  d^{10} x \sqrt{-g} e^{-2
\varphi} \Big(R + 4 |\partial \varphi|^2 - {1 \over 2} |H|^2\Big)\
, \cr T_{R-R} & = - {1 \over 2} d^{10} x \sqrt{-g} \sum_{n=0,1,2}\
|G_{2n}|^2 = - {1 \over 2} \sum_{n=0,1,2}\ G_{2n} \wedge \star
G_{2n}\ ,}} and $L_{CS}$ are the Chern-Simons terms in \iiadual ,
such that $S_{10} = {1 \over 2 \kappa_{10}^2} \int L_{10}$.
$T_{NS-NS}$ are the kinetic terms for the NSNS-fields and
$T_{R-R}$ will become those for the RR-fields after integrating
out the corresponding $A$-fields.

Plugging the decomposition of the $G$s and $A$s into \ldecomp\
results in the following expressions
\eqn\trdecomp{T_{R-R} = - {1
\over 2} G_0 \wedge \star G_0 - {1 \over 2} \sum_{n=1,2}\
\sum_{k=0,1,2} G_{2n}^{(k)} \wedge \star G_{2n}^{(k)}}
and
\eqn\lcsdecomp{\eqalign{ L_{CS} = & G_2^{(2)} \Big( {1 \over 2}
G_4^{(0)} (B^{(0)})^2 - {1 \over 3} G_2^{(0)} (B^{(0)})^3 + {1
\over 8} G_0 (B^{(0)})^4 + d^{(0)} A_7^{(0)} + B^{(0)} d^{(0)}
A_5^{(0)} \Big) \cr & + G_4^{(2)} \Big( - G_4^{(0)} B^{(0)} + {1
\over 2} G_2^{(0)} (B^{(0)})^2 - {1 \over 6} G_0 (B^{(0)})^3 -
d^{(0)} A_5^{(0)} \Big) \cr & - d^{(0)} A_9^{(2)} G_0 - d^{(1)}
A_9^{(1)} G_0 \cr & + \Big( d^{(0)} A_7^{(2)} + d^{(1)} A_7^{(1)}
\Big) \Big( G_2^{(0)} - G_0 B^{(0)} \Big) + d^{(0)} A_7^{(1)}
\Big( G_2^{(1)} - G_0 B^{(1)} \Big) \cr & + \Big( d^{(0)}
A_5^{(2)} + d^{(1)} A_5^{(1)} \Big) \Big( - G_4^{(0)} + G_2^{(0)}
B^{(0)} - {1 \over 2} G_0 (B^{(0)})^2 \Big) \cr & + d^{(0)}
A_5^{(1)} \Big( -G_4^{(1)} + G_2^{(1)} B^{(0)} + G_2^{(0)} B^{(1)}
- G_0 B^{(0)} B^{(1)} \Big) \cr &  + d^{(1)} A_7^{(0)} G_2^{(1)} -
d^{(1)} A_5^{(0)} \Big( G_4^{(1)} -G_2^{(1)} B^{(0)} \Big) \cr & +
B^{(2)} X_{(1)} + (B^{(1)})^2 X_{(2)} + B^{(1)} X_{(3)} + B^{(0)}
X_{(4)}\ ,}}
where
\eqn\x{\eqalign{ X_{(1)} = & - {1 \over 2} \Big( G_4^{(0)} -
G_2^{(0)} B^{(0)} + {1 \over 2} G_0 (B^{(0)})^2 \Big)^2 + \Big(
G_2^{(0)} - G_0 B^{(0)} \Big) d^{(0)} A_5^{(0)} - G_0 d^{(0)}
A_7^{(0)}\ , \cr X_{(2)} = & {1 \over 2} G_2^{(0)} G_4^{(0)} - {1
\over 2} (G_2^{(0)})^2 B^{(0)} - {1 \over 2} G_0 G_4^{(0)} B^{(0)}
+ {3 \over 2} G_0 G_2^{(0)} (B^{(0)})^2 \cr & - {1 \over 2} G_0^2
(B^{(0)})^3 - {1 \over 2} G_0 d^{(0)} A_5^{(0)}\ , \cr X_{(3)} = &
- \Big( G_4^{(0)} - G_2^{(0)} B^{(0)} + {1 \over 2} G_0
(B^{(0)})^2 \Big) \Big( G_4^{(1)} - G_2^{(1)} B^{(0)} \Big) - G_0
d^{(1)} A_7^{(0)} \cr & + \Big( G_2^{(0)} - G_0 B^{(0)} \Big)
d^{(1)} A_5^{(0)} + G_2^{(1)} d^{(0)} A_5^{(0)}\ , \cr X_{(4)} = &
- {1 \over 2} G_4^{(1)} \Big( G_4^{(1)} - G_2^{(1)} B^{(0)} \Big)
- {1 \over 6} (G_2^{(1)})^2 (B^{(0)})^2\ ,}}
which only contain $A^{(0)}$s.

Integrating out $A_5^{(2)}, A_7^{(2)}$ and $A_9^{(2)}$ (occurring
in the third to fifth line of \lcsdecomp), i.e. those components
with two external indices, leads to a similar solution as in
\solng . To be more precise we get \eqn\solntwo{\eqalign{G_0 & =
{\rm independent~of~} x^a\ , \cr G_2^{(0)} & = d^{(0)} A_1^{(0)} +
G_0 B^{(0)}\ , \cr G_4^{(0)} & = d^{(0)} A_3^{(0)} + G_2^{(0)}
B^{(0)} - {1 \over 2} G_0 (B^{(0)})^2\ . }} With the help of
\solntwo\ the expressions $X_{(1)}, X_{(2)}$ and $X_{(3)}$ from
\x\ simplify and become \eqn\xsimple{\eqalign{X_{(1)} & = - {1
\over 2} (d^{(0)} A_3^{(0)})^2 + d^{(0)} A_1^{(0)} d^{(0)}
A_5^{(0)} - G_0 d^{(0)} A_7^{(0)}\ , \cr X_{(2)} & = - {1 \over 2}
G_0 d^{(0)} A_5^{(0)} + {1 \over 2} d^{(0)} A_1^{(0)} d^{(0)}
A_3^{(0)} + {3 \over 4} G_0 d^{(0)} A_1^{(0)} (B^{(0)})^2 + {1
\over 2} G_0^2 (B^{(0)})^2\ , \cr X_{(3)} & = - d^{(0)} A_3^{(0)}
\Big( G_4^{(1)} - G_2^{(1)} B^{(0)} \Big) + d^{(0)} A_1^{(0)}
d^{(1)} A_5^{(0)} - G_0 d^{(1)} A_7^{(0)} + G_2^{(1)} d^{(0)}
A_5^{(0)}\ . }}

Furthermore integrating out $A_5^{(1)}, A_7^{(1)}$ and $A_9^{(1)}$
(occurring in the third to sixth line of \lcsdecomp), i.e. those
components with one external index, leads to the following
constraints:
\eqn\constone{\eqalign{& G_0 = {\rm independent~of~}
x^\mu\ , \cr & d^{(0)} \Big( G_2^{(1)} - G_0 B^{(1)} + d^{(1)}
A_1^{(0)} \Big) = 0\ , \cr & d^{(0)} \Big( G_4^{(1)} - G_2^{(1)}
B^{(0)} - d^{(0)} A_1^{(0)} B^{(1)} + d^{(1)} A_3^{(0)} \Big) = 0\
,}}
where the first equation implies, that the mass parameter is
(locally) independent of the two external coordinates and in the
last two equations \solntwo\ has been used. They are locally
solved by \eqn\solnone{\eqalign{G_2^{(1)} & = d^{(0)} A_1^{(1)} -
d^{(1)} A_1^{(0)} + G_0 B^{(1)}\ , \cr G_4^{(1)} & = d^{(0)}
A_3^{(1)} - d^{(1)} A_3^{(0)} + G_2^{(1)} B^{(0)} + d^{(0)}
A_1^{(0)} B^{(1)}\ . }}

The next step is to integrate out components of the fields ${\bf
G}$ with two external indices, i.e. $G_2^{(2)}$ and $G_4^{(2)}$.
The corresponding variations of the action $S_{10}$ lead to the
following equations of motion (c.f. the first two lines of
\lcsdecomp):
\eqn\tilgtwo{ * G_2^{(2)} = {1
\over 2} G_4^{(0)} (B^{(0)})^2 - {1 \over 3} G_2^{(0)} (B^{(0)})^3
+ {1 \over 8} G_0 (B^{(0)})^4 + d^{(0)} A_7^{(0)} + B^{(0)}
d^{(0)} A_5^{(0)} } and \eqn\tilgfour{
* G_4^{(2)} = - G_4^{(0)} B^{(0)} + {1 \over 2} G_2^{(0)}
(B^{(0)})^2 - {1 \over 6} G_0 (B^{(0)})^3 - d^{(0)} A_5^{(0)}\ . }
Thus we are left with $G_0, A_1^{(0)}, A_1^{(1)}, A_3^{(0)},
A_3^{(1)}, A_5^{(0)}$ and $A_7^{(0)}$ as the independent fields in
the RR-sector. They comprise a complete basis of independent
RR-fields in the ten-dimensional space time. All other components,
e.g. those with two external legs, can be expressed as duals of
them.

Introducing the notation
\eqn\gsixdef{G_6^{(0)} = - * G_4^{(2)}, \quad G_8^{(0)} =
* G_2^{(2)}\ ,}
it is easily verified that the relations \solntwo , \tilgtwo ,
\tilgfour\ between $A_{2n-1}^{(0)}$ and $G_{2n}^{(0)}$ take the
compact form
\eqn\aviagform{d^{(0)} {\bf A}^{(0)} = e^{-B^{(0)}} \wedge {\bf
G}^{(0)}\ ,}
which is similar to \solga, but now involves only forms with no
external components and also contains forms of degree six and
eight:
\eqn\ga{{\bf G}^{(0)} = \sum_{n=0}^4 G_{2n}^{(0)}, \quad {\bf
A}^{(0)} = \sum_{n=1}^4 A_{2n-1}^{(0)}\ .}
In $D$-dimensional Minkowski space a $p$-form transforms under
Hodge duality according to
\eqn\starstar{*(*\omega_{p}) = (-1)^{p(D-p)+1} \omega_{p}\ .}
Therefore, we can rewrite \gsixdef\ as
\eqn\tilgviadualg{G_4^{(2)} = * G_6^{(0)}, \quad G_2^{(2)} = -
* G_8^{(0)}\ .}

Now we substitute \gsixdef\ and \tilgviadualg\ into the Lagrangian
\ldecomp. From $T_{R-R}$ we get: \eqn\tf{ T_{R-R} = - {1 \over 2}
G_2^{(2)} \wedge * G_2^{(2)} - {1 \over 2} G_4^{(2)} \wedge
* G_4^{(2)} = {1 \over 2} G_8^{(0)} \wedge * G_8^{(0)}
+ {1 \over 2} G_6^{(0)} \wedge * G_6^{(0)}\ .}  There is a further
contribution to the kinetic terms of $G_6^{(0)}$ and $G_8^{(0)}$,
coming from the Chern-Simons terms: \eqn\csfviag{ L_{CS} =
G_2^{(2)} \wedge * G_2^{(2)} + G_4^{(2)} \wedge * G_4^{(2)} +
\ldots = - G_8^{(0)} \wedge
* G_8^{(0)} - G_6^{(0)} \wedge * G_6^{(0)} + \ldots
\ , } where we have used \tilgtwo\ and \tilgfour\ in the first two
lines of \lcsdecomp. Note, that the sum of the two contributions
\tf\ and \csfviag\ leads to the right sign for the kinetic terms
of $G_6^{(0)}$ and $G_8^{(0)}$.

Altogether we end up with the following expression for the dual
action:\foot{We have only considered the bosonic part of the action
here, which is all we need for a derivation of the potential.
We assume that the procedure can be made supersymmetric.}
\eqn\iiadual{\eqalign{ S_{10} = & {1 \over 2 \kappa_{10}^2} \int
d^{10} x \sqrt{-g} \Big(e^{-2 \varphi}(R + 4 |\partial \varphi|^2
- {1 \over 2} |H|^2 ) - {1 \over 2} \sum_{n=0}^4 |G_{2n}^{(0)}|^2
- {1 \over 2} \sum_{n=1}^2 |G_{2n}^{(1)}|^2 \Big) \cr & + {1 \over
2 \kappa_{10}^2} \int \Big( B^{(2)} X_{(1)} + (B^{(1)})^2 X_{(2)}
+ B^{(1)} X_{(3)} + B^{(0)} X_{(4)} \cr & + d^{(1)} A_7^{(0)}
G_2^{(1)} - d^{(1)} A_5^{(0)}( G_4^{(1)} - G_2^{(1)} B^{(0)})
\Big) + \ldots\ , }} where $X_{(1)}, \ldots ,X_{(4)}$ are given in
\x\ respectively \xsimple , $G_{2n}^{(1)}$ are defined in
\solnone\ and $G_{2n}^{(0)}$ are given in \aviagform , which can
be inverted to\foot{In order to get the correct $G_0$-dependence
in this formula, one formally has to introduce $A_{-1}^{(0)}$ with
a 0-form ''field strength'' $d^{(0)} A_{-1}^{(0)}$.}
\eqn\aviagforminv{{\bf G}^{(0)} = d^{(0)} {\bf A}^{(0)} \wedge
e^{B^{(0)}}\ .}
%

%%%%%%%%%%%%%%%%%%%%%%%%%%%%%%%%%%%%%%%%%%%%%%%%%%%%%%%%%%%%%%

\newsec{(Super)Potential Induced By Background Ramond-Ramond Fields}

In this section we use the Type IIA action \iiadual\ to obtain the
two-dimensional effective potential induced by background
RR-fields in a Kaluza-Klein reduction on a Calabi-Yau fourfold.
Following a similar analysis in M-theory \HaackL, we show that the
result can be expressed in terms of the superpotential for chiral
moduli superfields proposed in \GVW ,
\eqn\wone{W(Z^\alpha) = {1 \over 2 \pi} \int_{Y_4} F_4 \wedge
\Omega\ ,}
where $\Omega$ is the unique (4,0)-form of $Y_4$, and for twisted
chiral moduli superfields \Gukov:
\eqn\wtwo{\eqalign{ \tilde W(t^A) & = {1 \over 2 \pi} \int_{Y_4}
e^{- i \CK} \wedge {\cal F} = \cr & = {1 \over 2 \pi} \int_{Y_4}
\Big( {1 \over 4!} F_0 \CK^4 + i {1 \over 3!} F_2 \wedge \CK^3 -
{1 \over 2!} F_4 \wedge \CK^2 - i F_6 \wedge \CK + F_8 \Big)\ ,}}
where $\CK$ is a complexified K\"ahler form on $Y_4$.

Before we derive the potential, however, let us first recall the
compactification on a Calabi-Yau fourfold $Y_4$ without background
fluxes \MHLouis\ and see how it is reproduced in the dual
formulation.

\subsec{Reduction Without Fluxes}

Starting point for the reduction is the action \iiadual , albeit
with $G_0=0$. We take the space-time to be of the form $\IR^{1,1}
\times Y_4$, where $x^\mu$ are the coordinates of $\IR^{1,1}$ and
$x^a$ are those on the Calabi-Yau fourfold $Y_4$.\foot{However,
the determination of the massless spectrum and effective action in
$D=2$ makes use of the fact, that Calabi-Yau manifolds are complex
manifolds. We therefore introduce complex coordinates $\xi^i$ on
$Y_4$.} The spectrum of the $D=2$ theory is determined by the
deformations of the Calabi-Yau metric and the expansion of $B,
A_1^{(0)}, A_1^{(1)}, A_3^{(0)}, A_3^{(1)}, A_5^{(0)}$ and
$A_7^{(0)}$ in terms of the non-trivial forms of $Y_4$. The
deformations of the metric comprise $h^{1,1}$ real K\"ahler
deformations $M^A, A=1,\ldots, h^{1,1}$, and $h^{1,3}$ complex
deformations $Z^\alpha, \alpha=1,\ldots, h^{1,3}$, of the complex
structure. Since vectors contain no physical degree of freedom in
$D=2$ the modes $B^{(1)}, A_1^{(1)}$ and $A_3^{(1)}$ are
non-dynamical. Furthermore since there are no 1- and 7-forms on
$Y_4$ also $A_1^{(0)}$ and $A_7^{(0)}$ do not contribute any
massless mode in $D=2$. $B^{(0)}$ leads to $h^{1,1}$ real scalar
fields $a^A$, whereas expanding $A_3^{(0)}$ into the 3-forms and
$A_5^{(0)}$ into the 5-forms of $Y_4$, one immediately verifies,
that all $G_{2n}^{(0)}$ vanish, as do $G_2^{(1)}, X_{(1)},
X_{(2)}$ and $X_{(3)}$. The only contribution from the RR-sector
comes from \eqn\rrcontr{ S_{10}^{(RR)} = - {1 \over 4
\kappa_{10}^2} \int d^{10} x \sqrt{-g} |G_{4}^{(1)}|^2 - {1 \over
2 \kappa_{10}^2} \int \Big( {1 \over 2} B^{(0)} (G_4^{(1)})^2 +
d^{(1)} A_5^{(0)} G_4^{(1)} \Big)\ , } where
\eqn\gfourone{G_4^{(1)} = - d^{(1)} A_3^{(0)}\ ,} see \solnone .
In this case $A_5^{(0)}$ is just an auxiliary field, because it
does not have a kinetic term. However, in view of \gfourone , its
equation of motion $d^{(1)} G_4^{(1)} = 0$ is trivially fulfilled.
Thus the decomposition of $A_3^{(0)}$ into the (1,2)-forms of
$Y_4$ leads to $h^{1,2}$ complex scalars $N^I, I = 1,\ldots,
h^{1,2}$, with the same kinetic and interaction terms as in the
reduction of the usual action \MHLouis . The (1,1)-moduli reside
in twisted chiral multiplets, while all other scalars are members
of chiral multiplets. In the presence of both, chiral and twisted
chiral multiplets, the moduli space is in general not K\"ahler
anymore \GHR .

For simplicity let us discuss here only the case where the
$(2,1)$-moduli are frozen to some fixed value and refer to
\MHLouis\ for a discussion of the general case. Compactification
of the NS-part of the action \iiadual\ results in
\eqn\kinact{S_2 = {1 \over 2 \kappa_{10}^2} \int d^{2} x \sqrt{-g}
e^{-2 \varphi^{(2)}} \Big(R^{(2)} + 4 \partial_\mu \varphi^{(2)}
\partial^\mu \varphi^{(2)} - 2 g_{A \bar B} \partial_{\mu} t^A
\partial^{\mu} \bar t^B - 2 g_{\alpha \bar \beta} \partial_{\mu}
Z^{\alpha}
\partial^{\mu} \bar Z^{\bar \beta} \Big)\ , }
where the following definitions have been used\foot{Notice the
different normalization for the K\"ahler moduli as compared to
\MHLouis . This results in an additional factor of ${1 \over 2}$
in the definition of $g_{A \bar B}$ as compared to \refs{\HaackL,
\MHLouis, \HLouis}.} \refs{\GGWappear,\HLouis}:
\eqn\defs{\eqalign{ e^{-2 \varphi^{(2)}} & = e^{-2 \varphi} \CV \
, \cr t^A & = M^A + i a^A\ , \cr g_{A \bar B} & = {1 \over 4 \CV}
\int_{Y_4} e_A \wedge * e_B = -\partial_A \bar\partial_{{\bar B}}
\ln \CV \ , \cr g_{\alpha \bar{\beta}} & = - {\int_{Y_4}
 \Phi_{\alpha} \wedge \bar{\Phi}_{\bar{\beta}} \over
\int_{Y_4} \Omega \wedge \bar{\Omega}} = -\partial_{\alpha}
\bar\partial_{{\bar \beta}} \ln \Big( \int_{Y_4} \Omega \wedge
\bar{\Omega} \Big)\ .}}
Here $e_A$ denotes a basis for the (1,1)-forms of $Y_4$,
whereas $\Phi_\alpha$
a basis for its (3,1)-forms. $\CV$ is the volume of the fourfold,
which can be expressed in terms of the K\"ahler form $J=M^A e_A$
as
\eqn\vol{\CV = {1 \over 4!} \int_{Y_4} J\wedge J\wedge J\wedge J =
{1 \over 2^4 4!} d_{ABCD} (t^A + \bar t^A)(t^B + \bar t^B)(t^C +
\bar t^C)(t^D + \bar t^D)\ ,}
where  $d_{ABCD}$ are the classical intersection numbers
\eqn\intersect{d_{ABCD} = \int_{Y_4} e_A \wedge e_B \wedge e_C
\wedge e_D\ .}

The moduli space factorizes into chiral and twisted
chiral multiplets and is K\"ahler
despite the presence  of both kinds of multiplets.
The K\"ahler potential can be read off from \defs\ and is given by
\eqn\k{K = K_{3,1} + K_{1,1} = - \ln \Big( \int_{Y_4} \Omega
\wedge \bar{\Omega} \Big) - \ln \CV\ .}
Using this K\"ahler potential, the total kinetic action \kinact\
can be written in a manifestly supersymmetric form \GGWappear:
\eqn\generalkact{ S = \int d^2x \int d^2 \theta d^2 \bar \theta
E^{-1} e^{- 2V} e^{- K}\ ,}
where $V$ is a real superfield in ${\cal N}=(2,2)$ dilaton
supergravity. Its leading component can be identified with the
two-dimensional dilaton field.

This completes our brief review of the kinetic action for the moduli
fields in the effective $\CN=(2,2)$ two-dimensional supergravity.
We refer the reader to \refs{\GGWappear,\MHLouis,\HLouis}
for further details and a discussion of related aspects.

%%%%%%%%%%%%%%%%%%%%%%%%%%%%%%%%%%%%%%%%%%%%%%%%%%%%%%%%%%%%%%%%%%%%%

\subsec{Reduction With Fluxes}

We now come to the main issue of this paper, to the derivation of
the effective potential induced by RR-background fluxes in a
Calabi-Yau fourfold compactification of massive type IIA string theory.
The conditions for unbroken supersymmetry in this case have been
analyzed in \Gukov , where also a background flux for the NS
3-form field strength has been taken into account. In order to
preserve maximal symmetry of the two-dimensional space-time it has
to take the form $H_{\mu \nu m}= \epsilon_{\mu \nu} \partial_m f$,
where $f$ is a locally defined function of the internal
coordinates. Furthermore the space-time metric takes the form of a
warped product between a two-dimensional maximally symmetric space
and the internal Calabi-Yau manifold. It is shown in \Gukov\ that
the function $f$ is actually related via supersymmetry to the warp
factor $\Delta$ and the two-dimensional dilaton (see eq.\ (I.15)
of that paper). However, we do not consider possible effects of
the warp-factor on the potential here and in addition consider the
dilaton to be independent of the internal coordinates. Thus we
concentrate on the contributions from the kinetic terms of the
RR-fields, the Chern-Simons terms and some higher derivative
terms, assuming an unwarped metric in the derivation.\foot{The
warp-factor actually becomes constant in the large volume
limit \refs{\GKP, \GSS}. However, in contrast to the three-dimensional
case \HaackL , it is difficult to argue
that the contributions of the warp-factor are subleading in $\alpha'$,
as we will see momentarily. It would be nice to get a better
understanding of the role of the warp-factor.}

There are two higher derivative terms relevant for the derivation
of the potential. The first correction to \iiastand\ is given by
\refs{\VW, \DLM}
\eqn\firstcorr{ \delta S_1 = - T_1 \int B \wedge X_8\ ,}
where $T_1 = (2 \pi \alpha')^{-1}$ is the string tension and
\eqn\xeight{ X_8 = {1 \over 192\,(2 \pi)^4} \Big[ {\rm tr} R^4 -
{1 \over 4} ({\rm tr} R^2)^2 \Big]\ , \qquad \int_{Y_4} X_8\ =\
-{\chi \over 24}\ .}
In a Calabi-Yau fourfold compactification $\delta S_1$ leads to a
tadpole for $B^{(2)}$ if the fourfold has non-vanishing Euler
number \refs{\SVW, \DM}. This tadpole can be cancelled either by
introducing space-time filling strings into the vacuum, a
possibility that we do not pursue here, or by turning on
RR-background fluxes, as can be seen from the term $\sim B^{(2)}
X_{(1)}$ in \iiadual\ taken into account \xsimple . Thus if we
denote the background fluxes by\foot{To get a uniform notation we
also apply this definition to the 0-form background, i.e. $F_0
\equiv G_0$, compare the footnote at the end of section 2.}
\eqn\back{ F_{2n} = d^{(0)} A_{2n-1}^{(0)}\ ,}
the tadpole cancellation condition in the absence of space-time
filling strings is
\eqn\tadpole{{1 \over 4 \kappa^2_{10}} \int_{Y_4} \Big( 2 F_0
\wedge F_8 - 2 F_2 \wedge F_6 + F_4 \wedge F_4  \Big) = {T_1 \over
24} \chi\ ,}
which is a generalization of the three-dimensional tadpole
condition involving only the 4-form background \refs{\SVW, \DM,
\BB}. Since
\eqn\kappaten{ 2 \kappa^2_{10} = (2 \pi)^7 \alpha'^4\ ,}
we see from \tadpole , that the fluxes are of the order $F_{2n}
\sim {\cal O} (\alpha'^{n-1/2})$, which is in agreement with the 
usual flux quantization conditions.\foot{This can be clearly seen 
e.g.~from the formulas of appendix A in \KachruS.}
%, or to be more precise $F_{a_1
%\ldots a_{2n}} \sim {\cal O} (l_s^{2n-1}/l_Y^{2n})$, where $l_s$
%is the string length and $l_Y$ is the 'average radius' of $Y_4$.
%Thus in the decompactification limit $l_s/l_Y \rightarrow 0$ all
%fluxes vanish except $F_0$. This was to be expected, because our
%starting point was the massive IIA theory, which contains the mass
%parameter $F_0$ even in ten dimensions.

The second correction to \iiastand , that contributes to the
effective potential in $D=2$, is \refs{\GS, \GVZ, \GZ, \GW, \FPSS,
\PZ, \ST, \AFMN, \KP, \RT}
\eqn\scndcorr{ \delta S_2 = - {b_1 \over 4 \pi \alpha'} \int
d^{10} x \sqrt{-g^{(10)}} E_8\ ,}
where $b_1^{-1}\equiv (2 \pi)^4 3^2 2^{13}$ (we use the
conventions of \AT ) and
\eqn\eeight{E_8 = {1 \over 2} \epsilon^{ABM_1 N_1 \ldots M_4 N_4}
\epsilon_{ABM_1' N_1' \ldots M_4' N_4'} R^{M_1' N_1'}\! _{M_1 N_1}
\ldots R^{M_4' N_4'}\! _{M_4 N_4}\ .}
$E_8$ is a 10-dimensional generalization of the Euler density.
Thus, in a fourfold compactification this term contributes to the
potential according to \AT\
\eqn\eeightpot{- {b_1 \over 4 \pi \alpha'} \int d^{10} x
\sqrt{-g^{(10)}} E_8 = {T_1 \over 24} \int d^2 x \sqrt{-g^{(2)}}
\chi + \ldots\ ,}
where the dots stand for a correction to the two-dimensional
Einstein-Hilbert term, that can be derived along the same lines as
in \HaackL\ and which leads to a renormalization of the
two-dimensional dilaton. However, this is not important for our
discussion of the potential. Furthermore, the coefficient of the
$E_8$-term in \scndcorr\ does not get any higher string loop
corrections \GVH , such that the coefficient of the Euler term in
the effective potential is exactly given by \eeightpot .

There is one further well known higher derivative term \refs{\GS,
\GVZ, \GZ, \GW, \FPSS, \PZ, \ST, \AFMN, \KP, \RT},
\eqn\thrdcorr{\delta S_3 = \int d^{10} x \sqrt{-g^{(10)}} \Big(
{\alpha'^3 \zeta(3) \over 3 \cdot  2^{12} \kappa_{10}^2} e^{-2
\varphi} + {b_1 \over 2 \pi \alpha'} \Big) J_0 \ ,}
where
\eqn\jzero{ J_0 = t^{M_1 N_1 \ldots M_4 N_4} t_{M_1' N_1' \ldots
M_4' N_4'} R^{M_1' N_1'}\! _{M_1 N_1} \ldots R^{M_4' N_4'}\! _{M_4
N_4} + {1 \over 4} E_8\ . }
The tensor $t$ is defined by
$t^{M_1 \ldots M_8} A_{M_1 M_2} \ldots A_{M_7 M_8}
  = 24 {\rm tr} A^4 - 6 ({\rm tr} A^2)^2$
for antisymmetric tensors $A$, i.e. it does not contain the
$\epsilon$-term \Schwarz . If we assume that the dilaton $\varphi$
does not depend on the internal coordinates, this term does not
contribute to the two-dimensional potential, as the integral of $J_0$ over a
Calabi-Yau manifold vanishes \refs{\GW, \FPSS, \FT}.

Unlike in the three-dimensional case we can not exclude that other
higher derivative corrections might contribute to the potential at
the same order of $\alpha'$ as the kinetic terms of the RR-fields.
The reason for this is the different $\alpha'$-order of the
various fluxes, that we have seen above.\foot{This is also the
reason, why the warp-factor might modify the potential even to the
order we are working at.} This implies, for example, that a term
$\sim F_4^2 R^3$ would be of order ${\cal O}(\alpha'^6)$, while
the contribution of the kinetic term for $F_8$ is already of order
${\cal O}(\alpha'^7)$. Having said this caveat, however, we now
proceed calculating the effective potential, taking into account
only the terms appearing in \iiadual , \firstcorr\ and \scndcorr.
They give at least the leading contributions for each single flux.

Introducing the background fluxes \back\ into $S_{10} + \delta S_1
+ \delta S_2$ results in the following two-dimensional
action\foot{Certainly there are also corrections from higher
derivative terms to the kinetic terms displayed in \kinact , which
we do not calculate here.}
\eqn\fluxact{S_2 = S_2^{\rm (nf)} - \int d^{2} x \sqrt{-g^{(2)}}
V\ .}
Here $S_2^{\rm (nf)}$ is the action \kinact\ with no fluxes,
whereas the scalar potential $V$ is given by
\eqn\pot{\eqalign{V = {1 \over 4 \kappa_{10}^2} \Big[ \int_{Y_4}
d^8 \xi \sqrt{g^{(8)}} \Big( & |F_0|^2 + |F_2 + F_0 B|^2 + |F_4 +
F_2\wedge B + {1 \over 2} F_0 B^2|^2 \cr & + |F_6 + F_4 \wedge B +
{1 \over 2} F_2 \wedge B^2 + {1 \over 3!} F_0 B^3|^2 \cr & + |F_8
+ F_6 \wedge B + {1 \over 2} F_4 \wedge B^2 + {1 \over 3!} F_2
\wedge B^3+ {1 \over 4!} F_0 B^4|^2 \Big) \cr - \int_{Y_4} \Big( 2
F_0 F_8 - & 2 F_2 \wedge F_6 + F_4 \wedge F_4 \Big) \Big]\ ,}}
Deriving this expression we used \aviagforminv , \tadpole, and
\eeightpot, and also skipped the superscript of $B^{(0)}$ to avoid
cluttering. To proceed further we use the relation $\int_{Y_4} d^8
\xi \sqrt{g^{(8)}} |\omega_p|^2 = \int_{Y_4} \omega_p \wedge \star
\omega_p$, valid for an arbitrary p-form $\omega_p$. With the help
of the formulas for the Hodge-dual of even degree forms on
Calabi-Yau fourfolds given in appendix I, the potential \pot\ can
be expressed as
\eqn\pottwo{\eqalign{V = {1 \over 4 \kappa_{10}^2} \CV^{-1} \Big[
& F_0^2 \CV^2 + {1 \over 36} \Big( \int_{Y_4} (F_2 + F_0 B) \wedge
J^3 \Big)^2 - {1 \over 2} \CV \int_{Y_4} (F_2+  F_0 B)^2 \wedge
J^2 \cr & - \CV \int_{Y_4} F_{2,2} \wedge F_{2,2} + \CV \int_{Y_4}
(F_{2,2} + F_2 \wedge B + {1 \over 2} F_0 B^2)^2 \cr & + {1 \over
4} \Big( g^{-1 A \bar B} \int_{Y_4} (F_4 + F_2 \wedge B + {1 \over
2} F_0 B^2) \wedge J \wedge e_A  \cr & \, \, \, \, \, \, \, \, \,
\, \, \, \, \, \, \, \, \, \, \, \, \, \, \, \, \, \, \, \times
\int_{Y_4} (F_4 + F_2 \wedge B + {1 \over 2} F_0 B^2) \wedge J
\wedge e_B \cr & \, \, \, \, \, \, \, \, \, \, \, \, \, \, \, -
\Big( \int_{Y_4}(F_4 + F_2 \wedge B + {1 \over 2} F_0 B^2) \wedge
J^2 \Big)^2 \Big) \cr & + {1 \over 4} g^{-1 A \bar B}
\int_{Y_4}(F_6 + F_4 \wedge B + {1 \over 2} F_2 \wedge B^2 + {1
\over 3!} F_0 B^3) \wedge e_A \cr & \, \, \, \, \, \, \, \, \, \,
\, \, \, \, \, \, \, \, \, \, \, \, \, \, \times \int_{Y_4}(F_6 +
F_4 \wedge B + {1 \over 2} F_2 \wedge B^2 + {1 \over 3!} F_0 B^3)
\wedge e_B \cr & + \Big( \int_{Y_4}(F_8 + F_6 \wedge B + {1 \over
2} F_4 \wedge B^2 + {1 \over 3!} F_2 \wedge B^3 + {1 \over 4!} F_0
B^4) \Big)^2 \cr & - 2 \CV \int_{Y_4} F_0 F_8 + 2 \CV \int_{Y_4}
F_2 \wedge F_6 \Big] \cr - {1 \over \kappa_{10}^2} \int_{Y_4} &
F_{3,1} \wedge F_{1,3}\ .}}

The last term in \pottwo\ is identical to the potential for the
complex structure moduli appearing in three dimensions \HaackL .
As in that case it can be further rewritten by using \AS
\eqn\domega{ D_\alpha\Omega = \partial_\alpha \Omega +
(\partial_\alpha K_{3,1})\, \Omega = \Phi_\alpha\ ,}
where $\Phi_\alpha$ is the basis of $H^{3,1}(Y_4)$ and $K_{3,1}$
is the K\"ahler potential for the $(3,1)$-moduli defined in \k .
With the help of \domega\ and \defs\ one derives
\eqn\rewrite{ - {1 \over \kappa_{10}^2} \int_{Y_4} F_{3,1} \wedge
F_{1,3} = e^{K_{3,1}} g^{-1 \alpha\bar \beta} D_\alpha W D_{\bar
\beta} \bar W \ ,}
where $W$ is precisely the chiral superpotential of \GVW , given
in \wone , if we set $\kappa_{10} = 2 \pi$.

The potential for the K\"ahler moduli is more involved than in the
three dimensional case, due to the additional background fluxes
and the complexification of the K\"ahler moduli. It is a bit
tedious but straightforward to verify that the first term in
\pottwo\ can be expressed as
\eqn\rewritetilde{{1 \over 4 \kappa_{10}^2} \CV^{-1} \Big[ \ldots
\Big] = {1 \over 16} e^{K_{1,1}} g^{-1 A \bar B} D_A \tilde W
D_{\bar B} \bar{\tilde W}\ ,}
where $K_{1,1}$ is the K\"ahler potential for the K\"ahler moduli,
defined in \k , $\tilde W$ is the twisted chiral superpotential of
\wtwo\ and the K\"ahler covariant derivative is defined as $D_A
\tilde W = \partial_A \tilde W + (\partial_A K_{1,1}) \tilde W$.
The formulas necessary for the derivation are collected in
appendix I.

Obviously, defining
\eqn\hatw{ \hat W = {1 \over 4} \tilde W\ ,}
the potential takes the form
\eqn\potthree{V = e^{K_{1,1}} g^{-1 A \bar B} D_A \hat W D_{\bar
B} \bar{\hat W} + e^{K_{3,1}} g^{-1 \alpha\bar \beta} D_\alpha W
D_{\bar \beta} \bar W \ ,}
i.e. it does not contain any terms $\sim |W|^2$ or $\sim |\hat W|^2$.
The same phenomenon occurred in the three dimensional case
for the potential of the complex structure moduli and it is
closely related to the situation of four dimensional type IIB
compactifications with 3-form fluxes.

%In the appendix II we support this observation studying chiral
%superfields and density projectors in $\CN=2$ dilaton
%supergravity. Namely, we find that the form of chiral density
%projectors also suggests that terms $\sim |W|^2$ and $\sim |\hat
%W|^2$ do not appear in the scalar potential.

%%%%%%%%%%%%%%%%%%%%%%%%%%%%%%%%%%%%%%%%%%%%%%%%%%%%%%%%%%%%%%%%%%

\subsec{(Non)-supersymmetric Vacua}

In the effective two-dimensional theory, the conditions for
unbroken supersymmetry do not allow supersymmetric vacua with
non-zero cosmological constant and require the following
conditions to be satisfied \Gukov:
\eqn\susycond{D_A \tilde W = 0, \quad D_\alpha W = 0, \quad \tilde
W = 0, \quad W = 0\ .}
This is similar to the situation discovered in three dimensions
\refs{\GVW, \HaackL} . Also as in the three-dimensional case, a 4-form flux
$F_4 \sim \bar \Omega$ breaks supersymmetry without introducing a
vacuum energy \refs{\GKP, \GVW}. This is due to the fact that $W
\neq 0$ but $V=0$ according to \rewrite . A further flux with this
property, $F_4 \sim J^2$, has been found in \BBtwo . It has a
generalization in the two-dimensional situation at hand. Using
\pottwo\ and the formulas from appendix I one shows that for $B=0$
an arbitrary combination of the following fluxes leads to a vanishing
potential:
\eqn\zeropot{F_8 = {F_0 \over 4!} J^4\ , \qquad F_6 = - {1 \over
3!} F_2 \wedge J^2 \quad {\rm with} \quad F_2 \sim J\ , \qquad F_4
\sim J^2\ .}
Generically these fluxes break supersymmetry, because of $\tilde W
\neq 0$. However, the combination
\eqn\susyffour{F_8 = {F_0 \over 4!} J^4 \quad {\rm and} \quad F_4
= {F_0 \over 3!} J^2}
leads to a vanishing $\tilde W$ for $B=0$ and therefore to unbroken
supersymmetry.

This shows a general feature of the fourfold models in type IIA
string theory: they admit a rich vacuum structure due to the many
possible Ramond-Ramond fluxes that one can introduce in the
background. In fact, the non-trivial dependence of the potential
on the geometric moduli of the Calabi-Yau fourfold generically
allows one to stabilize all of them, including the volume modulus.
This is in contrast to similar compactifications of M-theory and
F-theory \refs{\GVW,\HaackL,\GKP}, where the volume remains a flat
direction of the tree-level potential.

To see that this problem does not occur in the type IIA theory,
let us consider the simple case of a Calabi-Yau fourfold with
$h^{1,1}(Y_4)=1$.\foot{Examples of such manifolds can be found e.g.
in \refs{\cyk, \cys}.} Thus
there is a single K\"ahler modulus $t=M+ia$. Furthermore let us assume that the
(1,1)-form $e$ is normalized in such a way that the intesection number
$\int e \wedge e \wedge e \wedge e =1$, so that $\CV={1 \over 4!} M^4$.
Finally let us consider only nonzero 0-form and 8-form flux. In this case
the potential \pottwo\ becomes
\eqn\potv{V = {1 \over 4 \kappa_{10}^2} M^{-4}
\left( {F_0^2 \over 4!} (M^2 + a^2)^4 + 2 F_0 \Big(\int_{Y_4} F_8\Big) (a^4 - M^4)
+ 4! \Big(\int_{Y_4} F_8\Big)^2 \right)\ .}
%
%where the dots denote the potential for the complex structure moduli.
%In fact, it is instructive to focus on this part of the potential
%assuming\foot{One can easily check this assumption by simply
%comparing the number of moduli with the number of all possible
%fluxes, which generate the potential \potthree.} that all the other
%moduli can be stabilized for any values of $F_8$, $F_0$, and $\CV$.
The structure of the vacuum of \potv\ depends on the relative
sign of the fluxes $F_0$ and $\int F_8$.
If both have the same sign, then the
only minimum of the potential $V$ is at:
\eqn\minam{a=0\ , \quad M=\Big({4! \over F_0} \int_{Y_4} F_8\Big)^{1/4}\ ,
\quad \CV = {1 \over F_0} \int_{Y_4} F_8\ .}
In this minimum $V=0$, but supersymmetry is broken because of a
non-vanishing $\tilde W$, as was already discussed above. However, if
one also turns on 4-form flux, the values \minam\ still lead to a minimum
of the potential at $V=0$.
In this case it is possible to find a supersymmetric minimum,
if \susyffour\ and the remaining conditions of  \susycond\ are satisfied.

On the other hand, if $F_0$ and $\int F_8$ have opposite signs,
the minimum of \potv\ occurs at:
\eqn\minav{a=0\ , \quad M=\Big(-{4! \over F_0} \int_{Y_4} F_8\Big)^{1/4}\ ,
\quad \CV = -{1 \over F_0} \int_{Y_4} F_8\ .}
Substituting this back into \potv, one finds the value of
the potential at the minimum:
\eqn\minv{V_{min} = {1 \over \kappa_{10}^2} \vert F_0 \int_{Y_4} F_8 \vert\ .}
In particular, note that $V_{min} > 0$, and the minimum is
classically stable.

This simple example should generalize to generic Calabi-Yau fourfolds
and fluxes. Therefore, one can easily construct not only many supersymmetric
vacua with fixed geometric moduli, but also configurations
with a positive value of the scalar potential in the minimum.
Such configurations obviously can not be supersymmetric,
but perhaps can be useful in the quest for de Sitter vacua
in string theory \refs{\Hull,\MNunez,\ChamblinL,\BHM,\Silverstein,\SSV,\BerglundAJ,\KPV}.\foot{We 
thank Andy Strominger for stimulating discussions on these aspects.}
Unfortunately, Ramond-Ramond fluxes alone do not yield a dS$_2$ vacuum.
This is because the  effective two-dimensional action one obtains from
a four-fold compactification is in the string frame
rather than in the Einstein frame and one can not generate a
potential for the dilaton by turning on Ramond-Ramond fluxes only.
One might hope to stabilize the dilaton by also introducing
background NS-NS 3-form fluxes as in \GKP.\foot{However,
this would require to start with the massless type IIA theory, 
that depends on the NS-NS 2-form only via its field strength \AJ.}
It would be interesting to pursue this further.

%%%%%%%%%%%%%%%%%%%%%%%%%%%%%%%%%%%%%%%%%%%%%%%%%%%%%%%%%%%%%%%%%%%%

\newsec{Examples}

Rather than demonstrating \rewritetilde\ in the general case, it
is more instructive to consider a few specific examples, which
illustrate the basic idea, and also help us to gain some intuition
about the physical effects of different background fluxes.

\subsec{Reduction with 8-form Flux}

The simplest non-trivial example is when $F_8$ is the only
non-trivial flux in the background. Then, from the dual Type IIA
action \iiadual, we expect the following scalar potential:
\eqn\vforgeight{V = {1 \over 4 \kappa^2_{10}} \int_{Y_4} d \xi^8
\sqrt{g^{(8)}} \vert F_8 \vert^2 = {1 \over 4 \kappa^2_{10}}
\CV^{-1} \Big( \int_{Y_4} F_8 \Big)^2\ .}
The Chern-Simons terms are not important in this case;
they only give a tadpole cancellation condition:
$$
\chi(Y_4)=0\ .
$$

Let us now demonstrate that this scalar potential is consistent with the proposed
expression for the superpotential. Specifically, from \wone\ and
\wtwo\ we get:
\eqn\wforgeight{W=0\ , \quad \tilde W = {1 \over 2 \pi} \int_{Y_4}
F_8 \in \IZ\ .}
The integer constant $\tilde W$ is nothing but the overall
D0-brane charge ``at infinity''. Specifically, a D0-brane looks
like a supersymmetric kink, or a BPS domain wall in the effective
$\CN=(2,2)$ two-dimensional theory. Moreover, it is magnetically
charged with respect to the $F_8$, so as we go across this domain
wall, the background value of the $F_8$-flux jumps by one unit.
Hence, we can think of the $F_8$ (in fact, of any Ramond-Ramond
flux) as being induced by D-brane charge placed at a large
distance.

Motivated by the form of \rewritetilde, let us evaluate:
\eqn\calcforgeight{\eqalign{ g^{-1A \bar B} D_A \tilde W D_{\bar
B} \bar {\tilde W} & = g^{-1 A \bar B} (\partial_A K_{1,1})
(\partial_{\bar B} K_{1,1}) \vert \tilde W \vert^2 \cr & = 4 g^{-1
A \bar B}\, \CV^{-2}\, \CV_A \CV_B\, \vert \tilde W \vert^2 \cr
% & = 4 \CV^{-2} (- {1 \over 3} \CV \CV^{-1AB} \CV_A \CV_B
%+ {4 \over 3} \CV^2) \vert \tilde W \vert^2 \cr & = 4 \CV^{-2} (-
%{1 \over 3} \CV M^A \CV_A + {4 \over 3} \CV^2) \vert \tilde W
%\vert^2 \cr & = 4 \CV^{-2} (- {1 \over 3} \CV^2 + {4 \over 3}
%\CV^2) \vert \tilde W \vert^2
& = 4 \CV^{-1} M^A \CV_A \vert \tilde W \vert^2 = 4 \vert \tilde W
\vert^2\ . }} Here we used some formulas from appendix I. This
result is in complete agreement with \vforgeight (if we use
$\kappa_{10} = 2 \pi$):
\eqn\vforwgeight{V = {1 \over 16} e^{K_{1,1}} g^{-1 A \bar B} D_A
\tilde W D_{\bar B} \bar {\tilde W} = {1 \over 4} \CV^{-1} \vert
\tilde W \vert^2\ .}
Notice, due to an inverse volume factor $\CV^{-1}$,
the scalar potential $V$ induced by the 8-form flux
has a minimum at large volume
$$
\CV \to \infty\ .
$$

%%%%%%%%%%%%%%%%%%%%%%%%%%%%%%%%%%%%%%%%%%%%%%%%%%%%%%%%%%%%%%

\subsec{Reduction with 0-form Flux}

Another extreme example is when 0-form flux is the only
non-vanishing R-R field in the background. Since $F_0$ plays the
role of the mass parameter in type IIA supergravity, this case
corresponds to compactification of massive type IIA supergravity
on a Calabi-Yau fourfold $Y_4$.

Even though this case looks simple, in fact, it is the most
difficult one. The reason is that when $F_0$ flux is non-zero, one
has to take into account terms with all the $G^{(0)}_{2n}$ fields
in the dual Type IIA action \iiadual, in order to reproduce the
right dependence of the potential on the moduli fields.
Specifically, $G^{(0)}_{2n}$ depends on $F_0$ via
$$
G^{(0)}_{2n} = {1 \over n!} F_0 B^n + \ldots \ .
$$

The superpotential \wtwo\ in this case reduces to
\eqn\wforgzero{\tilde W = {F_0 \over 2 \pi} {1 \over 4!}
\int_{Y_4} \CK^4 = {F_0 \over 2 \pi} {1 \over 4!} d_{ABCD} t^A t^B
t^C t^D = {F_0 \over 2 \pi} \mu\ , }
where we have introduced a holomorphic analog $\mu$ of the volume
$\CV$. Using the formulas from appendix I, it is easy to evaluate the
right hand side of \rewritetilde ,
\eqn\calcforgzero{\eqalign{ g^{-1 A \bar B} D_A \tilde W D_{\bar
B} \bar {\tilde W} & = g^{-1 A \bar B} (\partial_A \tilde W +
(\partial_A K_{1,1}) \tilde W ) (\bar
\partial_{B} \bar {\tilde W} + (\bar \partial_{B} K_{1,1}) \bar
{\tilde W}) \cr & = g^{-1 A \bar B} (4 {F_0 \over 2 \pi} \mu_A -
{2 \over \CV} \CV_A \mu {F_0 \over 2 \pi}) (4 {F_0 \over 2 \pi}
\bar \mu_{B} - {2 \over \CV} \CV_B \bar \mu {F_0 \over 2 \pi}) \cr
& = 4 \Big({F_0 \over 2 \pi}\Big)^2 g^{-1 A \bar B} \Big[ 4 \mu_A
\bar \mu_{B} + {1 \over \CV^2} \CV_A \CV_B \vert \mu \vert^2 - 2
\mu_A \CV_B {\bar \mu \over \CV} - 2 \CV_A \bar \mu_B {\mu \over
\CV} \Big] \cr & = 4 \Big({F_0 \over 2 \pi}\Big)^2 \Big[4 g^{-1 A
\bar B} \mu_A \bar \mu_{B} + \vert \mu \vert^2 - 2 M^A \mu_A \bar
\mu - 2 M^{B} \bar \mu_{B} \mu \Big]\ . }}

For simplicity let us make the further assumption that $B=0$ ({\it
i.e.} all $a^A=0$). Thus we have $\mu = \bar \mu = \CV$ and $\mu_A
= \bar \mu_A = \CV_A$ and \calcforgzero\ simplifies such that
\eqn\Vzero{V = {1 \over 16} e^{K_{1,1}} g^{-1 A \bar B} D_A \tilde
W D_{\bar B} \bar {\tilde W} = {1 \over 4} \Big({F_0 \over 2
\pi}\Big)^2 \CV = {1 \over 4} \CV^{-1} \vert \tilde W \vert^2\ .}

This result agrees with \pottwo\ for the case under consideration
and is very similar to \vforwgeight\ in the last subsection. Note,
however, that now the scalar potential has a minimum at small
volume:
$$
\CV \to 0\ .
$$

The moral is that $p$-form fluxes with small values of $p$ tend to
minimize the volume of the compactification space, whereas
$p$-form fluxes of high degree make the size of the
compactification manifold grow. This is a very general phenomenon,
known as attractor mechanism in higher-dimensional
supergravity solutions \refs{\FKS,\FK,\FGK}. 
Here, we can also arrange a very similar ``attractor'' behaviour,
balancing effects of various fluxes and thus fixing the volume modulus as
discussed at the end of the last section.
Explicit examples of supersymmetric vacua were obtained in this way
also in \Gukov. Fixing the volume seems to require $p$-form fluxes with
different values of $p$, whose contributions to the (super)potential
scale differently, compare e.g.\ the $J$-dependence of the various
contributions to the superpotential $\tilde W$ \wtwo .

%%%%%%%%%%%%%%%%%%%%%%%%%%%%%%%%%%%%%%%%%%%%%%%%%%%%%%%%%%%%%%

\subsec{Reduction with 4-form Flux}

In this subsection we would like to investigate the relation
of the two-dimensional potential for non-vanishing 4-form flux
to the corresponding three-dimensional potential. It has been
found in \HaackL , that the potential for M-theory on a fourfold is given
by\foot{This formula differs from the one given in \HaackL, because we are
interested here in the potential one gets without performing
a Weyl-rescaling in $D=3$. It is this theory that gives the
dilaton supergravity considered here upon reduction on a further circle.
Furthermore in contrast to \HaackL\ we have chosen $\kappa_{11}=(2 \pi)^{3/2}$
here in order to get $\kappa_{10}=2 \pi$.}
\eqn\pothree{- {1 \over 2 \pi}  \int d^3 x \sqrt{-g_{(M)}} \Big[ e^{K_{3,1}}
g^{-1 \alpha\bar \beta} D_\alpha W D_{\bar \beta} \bar W +
\CV^{-1}_{\rm (M)} \Big({1 \over 4} g^{-1 AB} \partial_A \tilde W_{(\rm r)} \partial_B
\tilde W_{(\rm r)} - \tilde W^2_{(\rm r)}  \Big)\Big]\ ,}
where $\tilde W_{(\rm r)}$ is a real version of $\tilde W$. To be more
precise
\eqn\tildewr{\tilde W_{(\rm r)} =  {1 \over 4} \int_{Y_4} {F_4 \over 2 \pi}
\wedge J_{\rm (M)} \wedge J_{\rm (M)}\ ,}
and the index at $g_{(M)}$,
$\CV_{\rm (M)}$ and $J_{\rm (M)}$ should remind at the fact,
that the quantities are defined using the M-theory metric\foot{Implicitly
also $g^{-1 AB}$ and  $\partial_A$ depend on it.}, which is related to
the type IIA string frame metric $g_{mn}^{(10)}$ via \wvar\
\eqn\miia{ds^2_{(11)} = g^{(11)}_{MN} dx^M dx^N = e^{- {2 \varphi \over 3}}
g_{mn}^{(10)} dx^m dx^n + e^{{4 \varphi \over 3}} (dx^{11} + C_m dx^m)^2\ .}
Note that we have the same fluxes in M-theory and type IIA-theory,
although in the M-theory compactification it is the eleven
dimensional analog of $dC_3$ from \iiastand\ which gets a flux and
in the type IIA case it is the $dA_3$ introduced in \solga . This
is due to the fact that we are only considering 4-form flux, in
which case the two different definitions of the flux indeed
coincide.

Reducing \pothree\ on the M-theory circle with radius $e^{{2 \varphi \over 3}}$
leads to the following contribution to the two-dimensional potential:
\eqn\potwo{- \int d^2 x \sqrt{-g^{(2)}} \Big[ e^{K_{3,1}}
g^{-1 \alpha\bar \beta} D_\alpha W D_{\bar \beta} \bar W +
e^{K_{1,1}} \Big( g^{-1 A \bar B} D_A \hat W
D_{\bar B} \bar{\hat W} \Big)\Big|_{\rm B=0} \Big]\ ,}
where $\hat W$ has been introduced in \hatw\ and the second term only involves the
real K\"ahler moduli. The contributions involving the moduli coming from the
$B$-field, (c.f. \pottwo ),
\eqn\bcontr{- \int d^2 x \sqrt{-g^{(2)}} \Big[
{1 \over 16} e^{K_{1,1}} \Big( g^{-1 A \bar B} \int_{Y_4} {F_4 \over 2 \pi}
\wedge B \wedge e_A \int_{Y_4} {F_4 \over 2 \pi} \wedge B \wedge e_B
+ (\int_{Y_4} {F_4 \over 2 \pi} \wedge B^2)^2 \Big) \Big]\ ,}
arise as follows. The three-dimensional action also contains
the terms \HaackL
\eqn\acthree{- {1 \over 2 \pi} \int d^3 x \sqrt{-g_{(M)}} \Big[
{1 \over 2 (2 \pi)^2} \CV_{(M)} g_{AB}
F^A_{mn} F^{B mn} +  {1 \over 4 \pi} \epsilon^{mnp}
\tilde W_{AB} A^A_m F^B_{np} \Big]\ ,}
where $\tilde W_{AB} = {1 \over 4} \int_{Y_4} {F_4 \over 2 \pi} \wedge e_A \wedge e_B$.
We reduce this to two dimensions using the metric
\eqn\miiathree{ds^2_{(3)} = g^{(3)}_{mn} dx^m dx^n = e^{- {2 \varphi \over 3}}
g_{\mu\nu}^{(2)} dx^\mu dx^\nu + e^{{4 \varphi \over 3}} (dx^{3} + C_\mu dx^\mu)^2\ ,}
c.f.\ \miia , which has the inverse
\eqn\ginverse{g^{(3)mn} = \left( \matrix{ e^{{2 \varphi \over 3}} g^{(2) \mu \nu}  &
-e^{{2 \varphi \over 3}} C^\mu  \cr
-e^{{2 \varphi \over 3}} C^\nu & e^{- {4 \varphi \over 3}}
+e^{{2 \varphi \over 3}} C_\rho C^\rho \cr }\right)\ .}
Furthermore we take the Ansatz \FS
\eqn\Athree{C^A_m = (A^A_\mu + C_\mu a^A, a^A)}
for the three-dimensional vectors. In this context the Ansatz for
the external components can be interpreted as a change from the
$C$-basis to the $A$-basis \aversuscbasis , because the $C^A_m$
arise from expanding the eleven-dimensional $C_3$ into the
(1,1)-forms of $Y_4$, while $a^A$ are the expansion coefficients
of $C_3$ into $e^A \wedge dx_3$, respectively of $B$ into $e^A$.
As $C_1 = A_1$ we see from \aversuscbasis\ that the $A^A_\mu$ come
from an expansion of $A_3$ into the (1,1)-forms of $Y_4$.

Introducing \miiathree , \ginverse\ and \Athree\ into \acthree\
leads to the following terms in the two-dimensional effective
action:
\eqn\actwo{\eqalign{- \int d^2 x \sqrt{-g^{(2)}} & \Big[ {1 \over 2 (2 \pi)^2} \CV g_{A \bar B}
(F^A_{\mu \nu} + a^A F^{\mu \nu}) (F^{B \mu \nu} + a^B F^{\mu \nu}) \cr
& + {1 \over 2 \pi} \epsilon^{\mu \nu} \tilde W_{AB}  \Big( a^A (F^B_{\mu \nu} + a^B F_{\mu \nu})
- {1 \over 2} a^A a^B F_{\mu \nu} \Big) \Big]\ ,}}
where $F_{\mu \nu}$ is the field strength of $C_\mu$ and $F^A_{\mu \nu}$ that
of $A^A_\mu$. In addition to \actwo\ there is a term
\eqn\kkvect{- {1 \over 8 (2 \pi)^2} \int d^2 x \sqrt{-g^{(2)}} \CV F_{\mu \nu}
F^{\mu \nu}}
from the reduction of the three-dimensional Einstein-Hilbert action. Integrating
out the vectorfields from \actwo\ and \kkvect , which do not have any dynamical
degrees of freedom in two dimensions, leads exactly to the part of
the potential involving the B-field moduli \bcontr , if one uses the relation
$B=a^A e_A$. Furthermore we notice that the equation of motion for $F_{\mu \nu}$ yields
\eqn\vectansatz{F_{\mu \nu} = 4 \pi \epsilon_{\mu \nu} \CV^{-1} \Big( \int_{Y_4}
{F_4 \over 2 \pi} \wedge B^2 \Big)\ ,}
which amounts to a non-trivial fibration of the M-theory circle over the
two-dimensional space-time.

%%%%%%%%%%%%%%%%%%%%%%%%%%%%%%%%%%%%%%%%%%%%%
\bigskip
\centerline{\bf Acknowledgments} We thank M.~Berg, R.~Kallosh,
J.~Louis, A.~Strominger, and E.~Witten for useful discussions.
This research was partially conducted during the period S.G. served
as a Clay Mathematics Institute Long-Term Prize Fellow. The work of
S.G.~is also supported in part by grant RFBR No. 01-02-17488, and
the Russian President's grant No. 00-15-99296. M.H.~would like to
thank the University of Princeton and especially I.~Klebanov for
hospitality at the beginning of the work. Moreover M.H.~thanks
the DFG for financial support and his work was supported in part by
I.N.F.N., by the EC contract HPRN-CT-2000-00122, by the EC contract
HPRN-CT-2000-00148, by the INTAS contract 99-0-590 and by the MURST-COFIN
contract 2001-025492.

%%%%%%%%%%%%%%%%%%%%%%%%%%%%%%%%%%%%%%%%%%%%%%%%%%%%%%%%%%%%%%%%%%%%%%%%
\vfil
\eject
\appendix{I}{Some Conventions and Useful Formulas}

In this appendix we summarize some conventions and useful
relations used throughout the paper. We tried to make our
conventions as close as possible to the existing notations in the
literature, yet still consistent:

${~~~~~~~~}$ $Y_4$ ${~~~~~~~~~~~~~~~~~~~}$ Calabi-Yau fourfold,

${~~~~~~~~}$ $\Omega$ ${~~~~~~~~~~~~~~~~~~~~}$ covariantly
constant holomorphic 4-form on $Y_4$,

${~~~~~~~~}$ $e_A \in H^2(Y_4,\IR)$ ${~~}$ a basis of 2-forms on
$Y_4$,

${~~~~~~~~}$ $J = M^A e_A$ ${~~~~~~~~}$ K\"ahler form on $Y_4$,

${~~~~~~~~}$ $B^{(0)} = a^A e_A$ ${~~~~~~}$ internal part of the
NS-NS 2-form field,

${~~~~~~~~}$ $\CK = J + i B^{(0)}$ ${~~~~}$ complexified K\"ahler
form on $Y_4$,

${~~~~~~~~}$ $t^A = M^A + i a^A$ ${~~}$ complexified K\"ahler
moduli of $Y_4$,

${~~~~~~~~}$ $Z^\alpha$ ${~~~~~~~~~~~~~~~~~~~}$ complex structure
moduli of $Y_4$,

${~~~~~~~~}$ $H = dB$ ${~~~~~~~~~~~~}$ NS-NS 3-form field
strength,

${~~~~~~~~}$ $G_{2n}$ ${~~~~~~~~~~~~~~~~~}$ Ramond-Ramond
$2n$-form field strength,

%${~~~~~~~~}$ $B_{01}$, ${~~~~~~~~~~~}$ external part of the NS-NS 2-form field,

${~~~~~~~~}$ $\varphi$ ${~~~~~~~~~~~~~~~~~~~~}$ ten-dimensional
dilaton field,

${~~~~~~~~}$ $\CV$ ${~~~~~~~~~~~~~~~~~~~~}$ volume of $Y_4$,
measured in the string frame metric.

We also define
\eqn\volab{\eqalign{ \CV & = {1 \over 4!} \int_{Y_4} J \wedge J
\wedge J \wedge J = {1 \over 2^4 4!} d_{ABCD} (t^A + \bar t^A)
(t^B + \bar t^B) (t^C + \bar t^C) (t^D + \bar t^D) \cr \CV_A & =
{1 \over 4!} \int_{Y_4} e_A \wedge J \wedge J \wedge J = {1 \over
4!} d_{ABCD} M^B M^C M^D = {1 \over 2} \partial_A \CV = {1 \over
2} \bar \partial_{\bar A} \CV \cr \CV_{AB} & = {1 \over 4!}
\int_{Y_4} e_A \wedge e_B \wedge J \wedge J = {1 \over 4!}
d_{ABCD} M^C M^D = {1 \over 3} \partial_A
\partial_B \CV\ ,}}
where $d_{ABCD}$ are the classical intersection numbers
\intersect\ . These quantities satisfy the following identities:
\eqn\volab{\eqalign{ \CV_A M^A & = \CV\ , \cr \CV_{AB} M^B & =
\CV_A\  , \cr g^{-1 A \bar B} \CV_B & = \CV M^A\ , \cr g^{-1 A
\bar B} & = - {1 \over 3} \CV \CV^{-1AB} + {4 \over 3} M^A M^B\
,}}
where the indices are raised with $\delta^{AB}$.

Below we list some formulas for the Hodge duals of differential
forms of even degree on $Y_4$:
\eqn\formduals{\eqalign{\star \omega_{0,0} = & {1 \over 4!}
\omega_{0,0} J^4\ , \qquad \star \omega_{4,4} = \CV^{-1}
\int_{Y_4} \omega_{4,4}\ , \cr \star \omega_{1,1} = & {1 \over 36}
\CV^{-1} \Big( \int_{Y_4} \omega_{1,1} \wedge J^3\Big) J^3 - {1
\over 2} \omega_{1,1} \wedge J \wedge J\ , \cr \star \omega_{3,3}
= & {1 \over 4} \CV^{-1} g^{-1 A \bar B} \Big(\int_{Y_4}
\omega_{3,3} \wedge e_B\Big) e_A \ , \cr \star \omega_{4,0} = &
\omega_{4,0}\ , \qquad \star \omega_{3,1} = - \omega_{3,1}\ ,
\qquad \star \omega_{1,3} = - \omega_{1,3}\ , \qquad \star
\omega_{0,4} = \omega_{0,4}\ , \cr \star \omega_{2,2} = &
\omega_{2,2} + {1 \over 4} \CV^{-1} \Big( g^{-1 A \bar B}
\Big(\int_{Y_4} \omega_{2,2} \wedge J \wedge e_A\Big) J \wedge e_B
- \Big(\int_{Y_4} \omega_{2,2} \wedge J^2 \Big) J^2 \Big)\ ,}}
which can be verified directly from the definition of the Hodge
star on Calabi-Yau fourfolds
\eqn\hstar{\eqalign{\star \omega_{p,q} = & {1 \over p! q! (4-p)!
(4-q)!}\, \, \omega_{i_1 \ldots i_p \bar \imath_1 \ldots \bar
\imath_q} \epsilon^{i_1 \ldots i_p}\, \! _{\bar \jmath_{p+1}
\ldots \bar \jmath_D} \epsilon^{\bar \imath_1 \ldots \bar
\imath_q}\, \! _{j_{q+1} \ldots j_D} \cr & \times d \xi^{j_{q+1}}
\wedge \ldots \wedge d \xi^{j_D} \wedge d \bar{\xi}^{\bar
\jmath_{p+1}} \wedge \ldots \wedge d \bar{\xi}^{\bar \jmath_{D}} \
.}}
%
%$$
%* \omega_2 = - {1 \over 2} \omega_2 \wedge J \wedge J
%+ {2 \over 3} {\int \omega_2 \wedge J \wedge J \wedge J \over \int
%J \wedge J \wedge J \wedge J} J \wedge J \wedge J
%$$
%
%$$
%G_4 = G^{(4,0)} + G^{(3,1)} + G^{(2,2)} + G^{(1,3)} + G^{(0,4)}
%$$
%
%$$
%G^{(2,2)} = G^{(2,2)}_0 + J \wedge G^{(1,1)}_0 + J \wedge J \wedge G^{(0,0)}_0
%$$
%
%$$
%* G_4 =  G_4 - 2 G^{(3,1)} - 2 G^{(1,3)} - 2 J \wedge G^{(1,1)}_0
%$$
%
Let us also introduce ``holomorphic analogs'' of $\CV$ and
$\CV_A$:
$$
\mu = {1 \over 4!} d_{ABCD} t^A t^B t^C t^D
$$
and
$$
\mu_A = {1 \over 4!} d_{ABCD} t^B t^C t^D, \quad \bar \mu_A = {1
\over 4!} d_{ABCD} \bar t^B \bar t^C \bar t^D\ .
$$

\vfil
\eject
%%%%%%%%%%%%%%%%%%%%%%%%%%%%%%%%%%%%%%%%%%%%%%%%%%%%%%%%%%%%%%%%%%%%%
\appendix{II}{Chiral Density Projectors In $\CN=(2,2)$ Dilaton Supergravity}

In this appendix we find the chiral and twisted chiral density
projectors in the $\CN=(2,2)$ dilaton supergravity constructed in
\GGWappear. As it was noticed in \GGWappear, unlike
in gauged $\CN=(2,2)$ supergravity \GWmeasure, here the chiral
projectors do not simply follow from the full superspace projector
since $\hat \nb^2 \CL$ is not a chiral superfield (for an
arbitrary $\CL$). Therefore, in order to solve the problem, we
will have to study chiral superfields in the new $\CN=(2,2)$
dilaton supergravity.

Let us first recall the supergravity algebra of the ``old''
$U(1)_A \otimes U(1)_V$ gauged dilaton supergravity \GGW:
$$
\{ \nabla_{+}, \nabla_{+}\} = 0\quad , \quad \{ \nabla_{-}, \nabla_{-}\} = 0\ ,
$$
\eqn\sugra{ \{ \nabla_{+}, \nabla_{\dot +}\} = i \nabla_{\pp}\quad ,
\quad \{ \nabla_{-}, \nabla_{\dot -}\} = i \nabla_{\mm}\ , }
$$
\{ \nabla_{+}, \nabla_{-} \} =
- {1 \over 2} \bar R (\CX - i \CY')\quad ,
\quad \{ \nabla_{+}, \nabla_{\dot -}\} =
- {1 \over 2} \bar F (\CX - i \CY)\ .
$$
Here, we follow the notations of \GGWappear.
In particular, the Lorentz
generators, $\CX$, and R-symmetry generators, $\CY$ and $\CY'$,
act on the covariant derivative $\nabla_{\a}$ as follows:
$$
[ \CX, \nabla_{\pm} ] = \pm \half \nabla_{\pm} \quad , \quad
[ \CX, \nabla_{\dot \pm} ] = \pm \half \nabla_{\dot \pm}\ ,
$$
\eqn\ycommut{[ \CY, \nabla_{\pm} ] = - {i \over 2} \nabla_{\pm}
\quad , \quad [ \CY, \nabla_{\dot \pm} ] = + {i \over 2}
\nabla_{\dot \pm}\ , }
$$
[ \CY', \nabla_{\pm} ] = \mp {i \over 2} \nabla_{\pm}
\quad , \quad [ \CY', \nabla_{\dot \pm} ] = \pm {i \over 2}
\nabla_{\dot \pm}\ .
$$

De-gauging the $U(1)_A \otimes U(1)_V$ R-symmetry:
\eqn\newderiv{\nabla_{\a} \to \hat \nabla_{\a} + \la_{\a} \CY +
\tilde \la_{\a} \CY'} gives the ``new'' dilaton supergravity
algebra:
$$
\{ \hat \nabla_{+}, \hat \nabla_{+}\} = i (\la_{+} + \tilde
\la_{+}) \hat \nabla_{+} \quad , \quad \{ \hat \nabla_{-}, \hat
\nabla_{-}\} = i (\la_{-} - \tilde \la_{-}) \hat \nabla_{-}\ ,
$$
$$
\{ \hat \nabla_{+}, \hat \nabla_{-} \} = - {1 \over 2} \bar R \CX
+ {i \over 2} (\la_{-} + \tilde \la_{-}) \hat \nabla_{+} + {i
\over 2} (\la_{+} - \tilde \la_{+}) \hat \nabla_{-}\ ,
$$
\eqn\newsugra{ \{ \hat \nabla_{+}, \hat \nabla_{\dot -} \} = - {1
\over 2} \bar F \CX + {i \over 2} (\la_{\dot -} + \tilde \la_{\dot
-}) \hat \nabla_{+} - {i \over 2} (\la_{+} - \tilde \la_{+}) \hat
\nabla_{\dot -}\ ,}
$$
\{ \hat \nabla_{+}, \hat \nabla_{\dot +}\} = i \hat \nabla_{\pp} +
{i \over 2} (\la_{\dot +} + \tilde \la_{\dot +}) \hat \nabla_{+} -
{i \over 2} (\la_{+} + \tilde \la_{+}) \hat \nabla_{\dot +}\ ,
$$
$$
\{ \hat \nabla_{-}, \hat \nabla_{\dot -}\} = i \hat \nabla_{\mm} +
{i \over 2} (\la_{\dot -} - \tilde \la_{\dot -}) \hat \nabla_{-} -
{i \over 2} (\la_{-} - \tilde \la_{-}) \hat \nabla_{\dot -}\ ,
$$
where, in order to solve the constraints (Bianchi identities),
one has:
\eqn\vla{\la_{+} = \tilde \la_{+} = i (\hat \nabla_{+} V) \quad ,
\quad \la_{-} = - \tilde \la_{-} = i (\hat \nabla_{-} V)\ ,}
$$
\la_{\dot +} = \tilde \la_{\dot +} = - i (\hat \nabla_{\dot +} V)
\quad , \quad
\la_{\dot -} = - \tilde \la_{\dot -} = - i (\hat \nabla_{\dot -} V)
$$
and
\eqn\vrf{ \bar R = 4 \hat \nabla_{-} \hat \nabla_{+} V
\quad , \quad \bar F = 4 \hat \nabla_{\dot -} \hat \nabla_{+} V\ .}
In the following we will also need the definition of their lowest
component fields \GGWappear:
\eqn\fgrh{ F \vert = G, \quad R \vert = H\ . }
In terms of the real unconstrained superfield $V$, we get the
following supergravity algebra:
$$
\{ \hat \nabla_{+}, \hat \nabla_{+}\} = - 2 (\hat \nabla_{+} V)
\hat \nabla_{+} \quad , \quad \{ \hat \nabla_{-}, \hat
\nabla_{-}\} = - 2 (\hat \nabla_{-} V) \hat \nabla_{-}\ ,
$$
\eqn\vsugra{ \{ \hat \nabla_{+}, \hat \nabla_{-} \} = - {1 \over
2} \bar R \CX \quad , \quad \{ \hat \nabla_{+}, \hat \nabla_{\dot
-} \} = - {1 \over 2} \bar F \CX \ ,}
$$
\{ \hat \nabla_{+}, \hat \nabla_{\dot +}\} = i \hat \nabla_{\pp} +
(\hat \nabla_{\dot +} V) \hat \nabla_{+} + (\hat \nabla_{+} V)
\hat \nabla_{\dot +}\ ,
$$
$$
\{ \hat \nabla_{-}, \hat \nabla_{\dot -}\} = i \hat \nabla_{\mm}
+ (\hat \nabla_{\dot -} V) \hat \nabla_{-}
+ (\hat \nabla_{-} V) \hat \nabla_{\dot -}\ .
$$

Let us consider the full superspace density projector:
\eqn\cproj{\eqalign{ & \int d^2x d^4 \th E^{-1} \CL = \cr & = \int
d^2x e^{-1} \big[ \nabla^2 + i \psi^{\dot -}_{\mm} \nabla_{+} - i
\psi^{\dot +}_{\pp} \nabla_{-} + (- {1 \over 2} \overline H -
\psi^{\dot -}_{\pp} \psi^{\dot +}_{\mm} + \psi^{\dot -}_{\mm}
\psi^{\dot +}_{\pp} )\big] \nb_{\dot +} \nb_{\dot -} \CL \vert =
\cr & = \int d^2x e^{-1} \big[ (\hat \nabla_{+} - (\hat \nabla_{+}
V)) (\hat \nabla_{-} - (\hat \nabla_{-} V)) + i \psi^{\dot
-}_{\mm} (\hat \nabla_{+} - (\hat \nabla_{+} V)) - i \psi^{\dot
+}_{\pp} (\hat \nabla_{-} - (\hat \nabla_{-} V)) + \cr & ~~~~~~~+
(- {1 \over 2} \overline H - \psi^{\dot -}_{\pp} \psi^{\dot
+}_{\mm} + \psi^{\dot -}_{\mm} \psi^{\dot +}_{\pp} )\big] \hat
\nb^2 \CL \vert = \cr & = \int d^2x e^{-1} \big[ \hat \nabla_{+}
\hat \nabla_{-} + i (\psi_{\mm}^{\dot -} - \la_{-}) \hat
\nabla_{+} - i (\psi_{\pp}^{\dot +} - \la_{+}) \hat \nabla_{-} +
\cr & ~~~~~~~+ \big( - {1 \over 4} \bar H - \psi_{\pp}^{\dot -}
\psi_{\mm}^{\dot +} + (\psi_{\mm}^{\dot -} - \la_{-})
(\psi_{\pp}^{\dot +} - \la_{+}) \big) \big] \hat \nabla_{\dot +}
\hat \nabla_{\dot -} \CL \vert\ . }}
Here, the first line is expressed in the
gauged covariant derivative $\nabla_{\a}$, whereas the last few
lines are expressed in terms of the covariant derivative $\hat
\nabla_{\a}$, which does not contain any gauge connection.

In the ``old'' gauged supergravity the measures are related as:
\eqn\measure{\int d^2x d^4 \th E^{-1} \CL = \int d^2x d^2 \th
\CE^{-1} \overline \nabla^2 \CL \vert\ .}
Furthermore, since for {\it any} superfield $\CL$ we have
$$
\nabla_{\dot +} \Big( \nabla_{\dot +} \nabla_{\dot -} \CL \Big) =
\nabla_{\dot -} \Big( \nabla_{\dot +} \nabla_{\dot -} \CL \Big) =
0\ ,
$$
we conclude that $\CL_c = \nb^2 \CL$ is a chiral superfield,
and we can interpret the expression in the first line of \cproj\
as a chiral density projector:
\eqn\aaa{\eqalign{ & \int d^2x d^4 \th E^{-1} \CL = \int d^2x
e^{-1} \big[ \nabla^2 + i \psi^{\dot -}_{\mm} \nabla_{+} - i
\psi^{\dot +}_{\pp} \nabla_{-} + \cr & + (- {1 \over 2} \overline
H - \psi^{\dot -}_{\pp} \psi^{\dot +}_{\mm} + \psi^{\dot -}_{\mm}
\psi^{\dot +}_{\pp} )\big] \nabla_{\dot +} \nabla_{\dot -} \CL
\vert = \int d^2x d^2 \th \CE^{-1} \overline \nabla^2 \CL \vert\
.}}
Hence, the chiral measure $\CE^{-1}$ is given, essentially,
by the expression in the square brackets:
\eqn\lcproj{\int d^2x d^2 \th \CE^{-1} \CL_c =
\int d^2x e^{-1} \big[ \nabla^2
+ i \psi^{\dot -}_{\mm} \nabla_{+}
- i \psi^{\dot +}_{\pp} \nabla_{-} + }
$$
+ (- {1 \over 2} \overline H
- \psi^{\dot -}_{\pp} \psi^{\dot +}_{\mm}
+ \psi^{\dot -}_{\mm} \psi^{\dot +}_{\pp} )\big] \CL_c \vert\ .
$$

This reasoning, however, can not be applied to the last lines in
\cproj\ in the new un-gauged supergravity. The reason is that
$\hat \nb^2 \CL$ is not a chiral superfield in this theory. This
can be easily seen, for example, if one hits this expression with
$\hat \nabla_{\dot +}$:
$$
\hat \nabla_{\dot +} \Big( \hat \nabla_{\dot +} \hat \nabla_{\dot
-} \CL\Big) = - (\hat \nabla_{\dot +} V) \hat \nabla_{\dot +} \hat
\nabla_{\dot -} \CL \ne 0\ .
$$

This example suggests how to modify $\hat \nb^2 \CL$ in order to
make it a chiral superfield. Let us define:
$$
\CL_c = e^V \hat \nabla_{\dot +} \hat \nabla_{\dot -} \CL\ .
$$
It is easy to check that $\CL_c$ defined this way is, in fact, a
chiral superfield in the un-gauged $\CN=(2,2)$ supergravity
theory:
\eqn\chka{\eqalign{ & \hat \nabla_{\dot +} \CL_c = \hat
\nabla_{\dot +} \left(e^V \hat \nabla_{\dot +} \hat \nabla_{\dot
-} \CL \right) = \cr & = e^V \Big( \hat \nabla_{\dot +} + (\hat
\nabla_{\dot +} V) \Big) \hat \nabla_{\dot +} \hat \nabla_{\dot -}
\CL = \cr & = e^V \Big( \hat \nabla_{\dot +}^2 + (\hat
\nabla_{\dot +} V) \hat \nabla_{\dot +} \Big) \hat \nabla_{\dot -}
\CL = ~~~~~ ({\rm{SUGRA~algebra}})\cr & = e^V \Big(- (\hat
\nabla_{\dot +} V) \hat \nabla_{\dot +} + (\hat \nabla_{\dot +} V)
\hat \nabla_{\dot +} \Big) \hat \nabla_{\dot -} \CL = 0\ . }}
In a similar way one can show:
\eqn\chkb{\eqalign{ & \hat \nabla_{\dot -} \CL_c = \hat
\nabla_{\dot -} \left(e^V \hat \nabla_{\dot +} \hat \nabla_{\dot
-} \CL \right) = ~~~~~ ({\rm{since}}~\CL~{\rm{is~scalar}}) \cr & =
- \hat \nabla_{\dot -} \left(e^V \hat \nabla_{\dot -} \hat
\nabla_{\dot +} \CL \right) = - e^V \Big( \hat \nabla_{\dot -} +
(\hat \nabla_{\dot -} V) \Big) \hat \nabla_{\dot -} \hat
\nabla_{\dot +} \CL = \cr & = - e^V \Big( \hat \nabla_{\dot -}^2 +
(\hat \nabla_{\dot -} V) \hat \nabla_{\dot -} \Big) \hat
\nabla_{\dot +} \CL = \cr & = e^V \Big((\hat \nabla_{\dot -} V)
\hat \nabla_{\dot -} - (\hat \nabla_{\dot -} V) \hat \nabla_{\dot
-} \Big) \hat \nabla_{\dot +} \CL = 0\ . }}
By analogy with the definition \aaa\ of the chiral density
projector in the ``old'' gauged supergravity, in the new
ungauged formalism we also want to find $\CE^{-1}$, such that
for $\CL_c = \exp (V) \hat \nb^2 \CL$ (with arbitrary superfield $\CL$),
the chiral superspace integral
$$
\int d^2x d^2 \th \CE^{-1} \CL_c =
\int d^2x d^2 \th \CE^{-1} e^V  \hat \nb^2 \CL
$$
could be written in terms of the full superspace integral.
Specifically, we get:
\eqn\aab{\eqalign{ & \int d^2x d^4 \th E^{-1} \CL = \cr & = \int
d^2x e^{-1} \big[ (\hat \nabla_{+} - (\hat \nabla_{+} V)) (\hat
\nabla_{-} - (\hat \nabla_{-} V)) + i \psi^{\dot -}_{\mm} (\hat
\nabla_{+} - (\hat \nabla_{+} V)) - i \psi^{\dot +}_{\pp} (\hat
\nabla_{-} - (\hat \nabla_{-} V)) + \cr & ~~~~~~~+ (- {1 \over 2}
\overline H - \psi^{\dot -}_{\pp} \psi^{\dot +}_{\mm} + \psi^{\dot
-}_{\mm} \psi^{\dot +}_{\pp} )\big] \hat \nb^2 \CL \vert = \cr & =
\int d^2x e^{-1} \big[ (\hat \nabla_{+} - (\hat \nabla_{+} V))
(\hat \nabla_{-} - (\hat \nabla_{-} V)) + i \psi^{\dot -}_{\mm}
(\hat \nabla_{+} - (\hat \nabla_{+} V)) - i \psi^{\dot +}_{\pp}
(\hat \nabla_{-} - (\hat \nabla_{-} V)) + \cr & ~~~~~~~+ (- {1
\over 2} \overline H - \psi^{\dot -}_{\pp} \psi^{\dot +}_{\mm} +
\psi^{\dot -}_{\mm} \psi^{\dot +}_{\pp} )\big] \cdot e^{-V} \cdot
e^{+V} \cdot \hat \nabla_{\dot +} \hat \nabla_{\dot -}\CL \vert =
\cr & = \int d^2x e^{-1} \exp(- \varphi) \big[ (\hat \nabla_{+} -
2 (\hat \nabla_{+} V)) (\hat \nabla_{-} - 2 (\hat \nabla_{-} V)) +
i \psi^{\dot -}_{\mm} (\hat \nabla_{+} - 2 (\hat \nabla_{+} V)) -
\cr & ~~~~~~~ - i \psi^{\dot +}_{\pp} (\hat \nabla_{-} - 2 (\hat
\nabla_{-} V)) + (- {1 \over 2} \overline H - \psi^{\dot -}_{\pp}
\psi^{\dot +}_{\mm} + \psi^{\dot -}_{\mm} \psi^{\dot +}_{\pp}
)\big] \cdot e^{+V} \cdot \hat \nabla_{\dot +} \hat \nabla_{\dot
-}\CL \vert\ , }}
where we pulled $e^{-V}$ all the way to the left. This resulted in
the $\exp(- \varphi)$ factor and in the difference in the
numerical coefficients.

Because $\CL_c =e^{+V} \hat \nabla_{\dot +} \hat \nabla_{\dot
+}\CL$ is chiral in the new supergravity, we can write a general
formula for any chiral $\CL_c$:
\eqn\aac{\eqalign{ & \int d^2x d^2
\th \CE^{-1} \CL_c = \cr & = \int d^2x e^{-1} \exp(- \varphi)
\big[ (\hat \nabla_{+} - 2 (\hat \nabla_{+} V)) (\hat \nabla_{-} -
2 (\hat \nabla_{-} V)) + i \psi^{\dot -}_{\mm} (\hat \nabla_{+} -
2 (\hat \nabla_{+} V)) - \cr & ~~~~~~~ - i \psi^{\dot +}_{\pp}
(\hat \nabla_{-} - 2 (\hat \nabla_{-} V)) + (- {1 \over 2}
\overline H - \psi^{\dot -}_{\pp} \psi^{\dot +}_{\mm} + \psi^{\dot
-}_{\mm} \psi^{\dot +}_{\pp} )\big] \CL_c \vert\ . }}
Using the commutation relations \newsugra\ in the new supergravity
algebra, we get:
\eqn\aad{\eqalign{ & \int d^2x d^2 \th \CE^{-1} \CL_c =
\cr & = \int d^2x e^{-1} \exp(- \varphi) \big[ \hat \nabla_{+}
\hat \nabla_{-} - 2 (\hat \nabla_{+} V) \hat \nabla_{-} + 2 (\hat
\nabla_{-} V) \hat \nabla_{+} - \cr & - 2 (\hat \nabla_{+} \hat
\nabla_{-} V) + 4 (\hat \nabla_{+} V)(\hat \nabla_{-} V) + i
\psi^{\dot -}_{\mm} \hat \nabla_{+} - 2 i \psi^{\dot -}_{\mm}
(\hat \nabla_{+} V) -\cr & - i \psi^{\dot +}_{\pp} \hat \nabla_{-}
+ 2 i \psi^{\dot +}_{\pp} (\hat \nabla_{-} V) + (- {1 \over 2}
\overline H - \psi^{\dot -}_{\pp} \psi^{\dot +}_{\mm} + \psi^{\dot
-}_{\mm} \psi^{\dot +}_{\pp} )\big] \CL_c \vert = \cr & = \int
d^2x e^{-1} \exp(- \varphi) \big[ \hat \nabla_{+} \hat \nabla_{-}
+ 2 i \la_+ \hat \nabla_{-} - 2 i \la_- \hat \nabla_{+} + {\bar R
\over 2} - 4 \la_+ \la_- + \cr & + i \psi^{\dot -}_{\mm} \hat
\nabla_{+} - 2 \psi^{\dot -}_{\mm} \la_+ - i \psi^{\dot +}_{\pp}
\hat \nabla_{-} + 2 \psi^{\dot +}_{\pp} \la_- + (- {1 \over 2}
\overline H - \psi^{\dot -}_{\pp} \psi^{\dot +}_{\mm} + \psi^{\dot
-}_{\mm} \psi^{\dot +}_{\pp} )\big] \CL_c \vert = \cr & = \int
d^2x e^{-1} \exp(- \varphi) \big[ \hat \nabla_{+} \hat \nabla_{-}
+ i (\psi^{\dot -}_{\mm} - 2 \la_-) \hat \nabla_{+} - i
(\psi^{\dot +}_{\pp} - 2 \la_+) \hat \nabla_{-}- \cr & -
\psi^{\dot -}_{\pp} \psi^{\dot +}_{\mm} + (\psi^{\dot -}_{\mm} - 2
\la_-) (\psi^{\dot +}_{\pp} - 2 \la_+) \big] \CL_c \vert\ . }}
In the last equality we have used \fgrh . Summarizing, we find the
following chiral density projector formula:
\eqn\aaf{\eqalign{ & \int d^2x d^2 \th \CE^{-1} \CL_c =
\int d^2x e^{-1} \exp(- \varphi) \big[ \hat \nabla_{+} \hat
\nabla_{-} + i (\psi^{\dot -}_{\mm} - 2 \la_-) \hat \nabla_{+} -
\cr & - i (\psi^{\dot +}_{\pp} - 2 \la_+) \hat \nabla_{-}
 - \psi^{\dot -}_{\pp} \psi^{\dot +}_{\mm}
+ (\psi^{\dot -}_{\mm} - 2 \la_-) (\psi^{\dot +}_{\pp} - 2 \la_+)
\big] \CL_c  \vert\ . }}

Note, unlike the ``old'' chiral density projector,
the right-hand side of this expression does not contain
an $\bar H$-term.
%It implies that the corresponding
%scalar potential has no $\vert W \vert^2$ term!
This means, however, that one runs into an obvious problem if one
tries to incorporate the superpotential in the conventional way,
i.e. via
\eqn\superpot{S = S_0 + \int d^2x d^2 \th \CE^{-1} W\ .}
To see this let us evaluate the following superspace action for
chiral superfields $\Phi_i$, c.f.~\generalkact\ and \GGWappear :
\eqn\aaf{\eqalign{ & S =
%\int d^2x d^4 \th E_0^{-1} e^{-K} + \int d^2x d^2 \th \CE^{-1} W = \cr &
\int d^2x d^4 \th E^{-1} \exp(-2V) \exp(-K)
+ \int d^2x d^2 \th \CE^{-1} W\ .
}}
The terms relevant for the potential are
$$
L_{\rm aux} \sim
e^{-2 \varphi - K} \big[ \vert \half H + i \bar A_i
{\partial K \over \partial \phi_i} \vert^2
- {\partial^2 K \over \partial \phi_i \partial \bar \phi_j} A_i \bar A_j\big]
- i {\partial W \over \partial \phi_i} A_i e^{- \varphi} + {\rm c.c.}\ .
$$
Integrating out the auxiliary field $H$, we get:
$$
L_{\rm aux} \sim e^{-2 \varphi - K}
\Big({\partial^2 K \over \partial \phi_i \partial \bar \phi_j} \Big)
A_i \bar A_j
- i \Big( {\partial W \over \partial \phi_i} \Big)
A_i e^{- \varphi} + {\rm c.c.}\ .
$$
Finally, integrating out $A_i$ we obtain:
\eqn\aag{ L
\sim e^{K} K^{-1}_{\phi_i \bar \phi_j} \Big( {\partial W \over
\partial \phi_i} \Big) \overline{\Big( {\partial W \over \partial
\bar \phi_j} \Big)}\ ,}
without the $\vert W \vert^2$-term and without any dependence on the
dilaton. However, the derivatives that enter \aag\ are the ordinary
partial derivatives, whereas we expect them to be replaced by
suitable covariant derivatives in the theory of gravity. At the moment
it is not clear to us, how to modify \aaf\ in order to resolve this problem.
%This
%seems to be a general problem, unless one uses a more general
%ansatz for the supergravity action, not restricting to the
%exponential dependence $\exp(-2V)$ on the real superfield $V$:
%$$
%\int d^2x d^4 \theta E^{-1} \CL (V, \Phi_i, \Sigma_j)\ .
%$$

Similarly, one can obtain the twisted chiral density projector. We
can define a twisted chiral field as (for arbitrary superfield
$\CL$):
$$
\CL_{tc} = e^V \hat \nabla_{\dot +} \hat \nabla_{-} \CL\ .
$$
Indeed, it satisfies:
$$
\hat \nabla_{\dot +} \CL_{tc} = \hat \nabla_{-}\CL_{tc} =0\ .
$$
The proof is very similar to \chka\ and \chkb:
\eqn\chkc{\eqalign{ &\hat \nabla_{\dot +}\CL_{tc} =\hat
\nabla_{\dot +} \left(e^V \hat \nabla_{\dot +} \hat \nabla_{-} \CL
\right) = \cr & = e^V \Big( \hat \nabla_{\dot +} + (\hat
\nabla_{\dot +} V) \Big) \hat \nabla_{\dot +} \hat \nabla_{-} \CL
= \cr &= e^V \Big( \hat \nabla_{\dot +}^2 + (\hat \nabla_{\dot +}
V) \hat \nabla_{\dot +} \Big) \hat \nabla_{-} \CL =0}}
and
\eqn\chkd{\eqalign{ &\hat \nabla_{-}\CL_{tc} =\hat \nabla_{-}
\left(e^V \hat \nabla_{\dot +} \hat \nabla_{-} \CL \right) = \cr &
= -\hat \nabla_{-} \left(e^V \hat \nabla_{-} \hat \nabla_{\dot +}
\CL \right) = -e^V \Big( \hat \nabla_{-} + (\hat \nabla_{-} V)
\Big) \hat \nabla_{-} \hat \nabla_{\dot +} \CL = \cr &= -e^V \Big(
\hat \nabla_{-}^2 + (\hat \nabla_{-} V) \hat \nabla_{-} \Big) \hat
\nabla_{\dot +} \CL =0\ . }}

The twisted chiral density projection formula in
the $U(1)_V$ gauged supergravity theory has
the following form, analogous to \cproj:
\eqn\tproj{\eqalign{
& \int d^2x d^4 \th E^{-1} \CL = \cr
& = \int d^2x e^{-1} \big[ \nabla_{\dot -} \nabla_{+} -
i \psi^{-}_{\mm} \nabla_{+} + i \psi^{\dot +}_{\pp} \nabla_{\dot -}
+ ( {1 \over 2} \overline G + \psi^{-}_{\pp} \psi^{\dot +}_{\mm} +
\psi^{\dot +}_{\pp} \psi^{-}_{\mm} )\big]
\nabla_{\dot +} \nabla_{-} \CL \vert \cr
& = \int d^2x e^{-1} \big[
(\hat \nabla_{\dot -} - (\hat \nabla_{\dot -} V))
(\hat \nabla_{+} - (\hat \nabla_{+} V))
- i \psi^{-}_{\mm}
(\hat \nabla_{+} - (\hat \nabla_{+} V)) + \cr
& ~~~~~~~ + i \psi^{\dot +}_{\pp}
(\hat \nabla_{\dot -} - (\hat \nabla_{\dot -} V))
+ ( {1 \over 2} \overline G + \psi^{-}_{\pp} \psi^{\dot +}_{\mm} +
\psi^{\dot +}_{\pp} \psi^{-}_{\mm} )\big]
\hat \nabla_{\dot +} \hat \nabla_{-} \CL \vert\ . }}
Using the same arguments as above, we obtain the twisted chiral
density projection formula in the dilaton $\CN=(2,2)$
supergravity, c.f.~\aab:
\eqn\bbb{\eqalign{ & \int d^2x d^4
\th E^{-1} \CL = \cr & = \int d^2x e^{-1} \big[ (\hat \nabla_{\dot
-} - (\hat \nabla_{\dot -} V)) (\hat \nabla_{+} - (\hat \nabla_{+}
V)) - i \psi^{-}_{\mm} (\hat \nabla_{+} - (\hat \nabla_{+} V)) + i
\psi^{\dot +}_{\pp} (\hat \nabla_{\dot -} - (\hat \nabla_{\dot -}
V)) + \cr & ~~~~~~~+ ( {1 \over 2} \overline G + \psi^{-}_{\pp}
\psi^{\dot +}_{\mm} + \psi^{\dot +}_{\pp} \psi^{-}_{\mm} )\big]
\hat \nabla_{\dot +} \hat \nabla_{-} \CL \vert = \cr & = \int d^2x
e^{-1} \big[ (\hat \nabla_{\dot -} - (\hat \nabla_{\dot -} V))
(\hat \nabla_{+} - (\hat \nabla_{+} V)) - i \psi^{-}_{\mm} (\hat
\nabla_{+} - (\hat \nabla_{+} V)) + i \psi^{\dot +}_{\pp} (\hat
\nabla_{\dot -} - (\hat \nabla_{\dot -} V)) + \cr & ~~~~~~~+ ( {1
\over 2} \overline G + \psi^{-}_{\pp} \psi^{\dot +}_{\mm} +
\psi^{\dot +}_{\pp} \psi^{-}_{\mm} )\big] \cdot e^{-V} \cdot
e^{+V} \cdot \hat \nabla_{\dot +} \hat \nabla_{-} \CL \vert = \cr
& = \int d^2x e^{-1} \exp(- \varphi) \big[ (\hat \nabla_{\dot -} -
2 (\hat \nabla_{\dot -} V)) (\hat \nabla_{+} - 2 (\hat \nabla_{+}
V)) - i \psi^{-}_{\mm} (\hat \nabla_{+} - 2 (\hat \nabla_{+} V)) +
\cr & ~~~~~~~+ i \psi^{\dot +}_{\pp} (\hat \nabla_{\dot -} - 2
(\hat \nabla_{\dot -} V)) + ( {1 \over 2} \overline G +
\psi^{-}_{\pp} \psi^{\dot +}_{\mm} + \psi^{\dot +}_{\pp}
\psi^{-}_{\mm} )\big] \cdot e^{+V} \cdot \hat \nabla_{\dot +} \hat
\nabla_{-} \CL \vert\ .}}
By replacing $e^V \hat \nabla_{\dot +} \hat \nabla_{-} \CL$ with
an arbitrary twisted chiral superfield and following the same
steps as in \aac\ and \aad , we arrive at
\eqn\bbf{\eqalign{& \int d^2x d^2 \th \tilde \CE^{-1} \CL_{tc} =
\int d^2x e^{-1} \exp(- \varphi) \big[ \hat \nabla_{\dot -} \hat
\nabla_{+} - i (\psi^{-}_{\mm} + 2 \la_{\dot -}) \hat \nabla_{+} +
\cr & + i (\psi^{\dot +}_{\pp} - 2 \la_+) \hat \nabla_{\dot -} +
\psi^{-}_{\pp} \psi^{\dot +}_{\mm} - (\psi^{-}_{\mm} + 2 \la_{\dot
-}) (\psi^{\dot +}_{\pp} - 2 \la_+) \big] \CL_{tc}
\vert\ .}}

\listrefs
\end